\documentclass[hyper,12pt,letterpaper]{JHEP3}
\usepackage{epic,eepic}
\usepackage{graphicx}

\title{Duality Symmetry and the Cardy Limit}

\preprint{\hepth{yymmnnn}\\IITM/PH/TH/2007/14\\TIFR/TH/07-35}
\author{Suresh Nampuri\footnote{Email:
suresh@theory.tifr.res.in}$^{~1}$,Prasanta K.Tripathy\footnote{Email:
prasanta@physics.iitm.ac.in}$^{~2}$,and Sandip P. Trivedi\footnote{Email:
sandip@theory.tifr.res.in}$^{~1}$\\
\it $^1$Department of Theoretical Physics\\
\it Tata Institute of Fundamental Research\\
\it Homi Bhabha Rd, Mumbai 400 005, India.\\

\it $^2$Department of Physics,\\Indian Institute of Technology Madras,\\Chennai 600 036, India.\\

}

\abstract{We study supersymmetric and non-supersymmetric extremal  black holes obtained in
 Type IIA string theory compactified on $K3\times T^2$, with duality group $O(6,22,\mathbb Z) \times SL(2,\mathbb Z)$.
In the Cardy limit an internal circle combines with the  $AdS_2$ component in the near horizon geometry
 to give a BTZ black hole whose entropy is given by the Cardy formula.
We study black holes carrying $D0-D4$  and $D0-D6$ brane charges.  We find, both in the
supersymmetric and non-supersymmetric cases,  that a generic set of charges cannot be brought to
 the Cardy limit using the duality symmetries.
 In the non-supersymmetric case, unlike the
supersymmetric one,  we find that when the charges are large,  a small fractional change in them
 always allows the charges to  be taken to the Cardy limit.
These results could  lead to a microscopic determination of the entropy
for extremal non-supersymmetric black holes, including rotating cases like
 the extreme Kerr black hole in four dimensions.

}


\begin{document}
\newenvironment{theorem}{\begin{quote} \footnotesize }{ \normalsize \end{quote}}

\newcommand{\BesselI}[3]{\hat I_{#1}\left( #2 \pi \sqrt{ #3 }\right)}

\renewcommand{\th}{\theta}
\renewcommand{\Im}{\mbox{Im}}
\renewcommand{\Re}{\mbox{Re}}

\newcommand{\Zar}[2]{Z\!\left[ {}^{#1}_{#2} \right]}

\newcommand{\ar}[2]{\left[ {}^{#1}_{#2} \right]}
\newcommand{\art}[2]{\left[ {}^{\frac{#1}{2}}_{\frac{#2}{2}} \right]}

\newcommand{\pa}{\partial}
\newcommand{\nn}{\nonumber}
\newcommand{\eps}{\epsilon}
\newcommand{\IR}{\mathbb{R}}
\newcommand{\IZ}{\mathbb{Z}}
\newcommand{\IX}{\mathbb{X}}
\newcommand{\Zint}{\mathbb{Z}}
\newcommand{\Nint}{\mathbb{N}}
\newcommand{\Tr}{\mbox{Tr}}
\newcommand{\sgn}{\mbox{sgn}}
\newcommand{\CC}{\cal{C}}

\newcommand{\apm}{\alpha'}
\def\d{\partial}

\def\x{\xi}
\def\r{\rho}
\def\f{\phi}
\def\s{\sigma}
\def\g{\gamma}
\def\t{\tau}
\def\a{\alpha}
\def\b{\beta}
\def\m{\mu}
\def\n{\nu}
\def\e{\epsilon}
\def\w{\omega}
\def\h{\eta}
\def\l{{\lambda}}
\def\la{{\lambda}}
\def\o{{\omega}}
\def\D{\Delta}
\def\G{\Gamma}
\def\L{I}
\def\zn{Z_N}
\def\wp{{\cal P}}
\def\vt#1#2#3{ {\vartheta[{#1 \atop  #2}](#3\vert \tau)} }
%

\def\CH{{\cal H}}
\def\CR{{\cal R}}
\def\CM{{\cal M}}
\def\CF{{\cal F}}
\def\CS{{\cal S}}
\def\CL{{\cal L}}
\def\CI{{\cal I}}
\def\CD{{\cal D}}
\def\CZ{{\cal Z}}
\def\CN{{\cal N}}
\def\O{{\cal O}}
\def\CP{{\cal P}}
\def\CQ{{\cal Q}}
\def\sN{\scriptscriptstyle N}
\def\tN{{\tilde N}}
\def\half{{\frac12}}
\def\ap{$\alpha '$}
\def\IC{\relax\hbox{$\inbar\kern-.3em{\rm C}$}}
\def\bZ{{\bf Z}}
\def\bI{{\bf I}}
\def\bM{{\bf M}}
\def\bC{{\bf C}}
\def\bN{{\bf N}}
\def\bR{{\bf R}}
\def\bT{{\bf T}}
\def\bS{{\bf S}}
\def\bK{{\bf K}}
\def\zb{\bar z}
\def\zt{{\tilde z}}
\def\yt{{\tilde y}}
\def\xt{{\tilde x}}
\def\ft{{\tilde f}}
\def\gt{{\tilde g}}
\def\tQ{{\tilde Q}}

\def\IC{{\bf C}}
\def\IP{{\bf P}}
\def\IQ{{\bf Q}}
\def\CN{{\cal N}}
\def\CQ{{\cal Q}}
\def\CX{{\cal X}}
\def\CZ{{\cal Z}}

\newcommand{\ir}[1]{\ensuremath{\boldsymbol{#1}}}
\def\bea{\begin{eqnarray}}
\def\eea{\end{eqnarray}}
\def\be{\begin{equation}}
\def\ee{\end{equation}}
\def\ba{\begin{align}}
\def\ea{\end{align}}
\def\bse{\begin{subequations}}
\def\ese{\end{subequations}}
\def\1F1{{}_1\!F_1}
\def\2F0{{}_2\!F_0}

\def\qeq{\, {? \atop ~} \hskip-4mm =}

\def\bP{$\bar{\rm P}$\,}

\def\slr{$SL(2,R)$}
\def\ve{\epsilon}

\def\jt{\tilde{j}}
\def\mt{\tilde{m}}
\def\Nt{\tilde{N}}
\def\ni{\noindent}
\def\ga{\gamma}
\def\Ga{\Gamma}
\def\bm{\bar{m}}
\def\bx{\bar{x}}
\def\by{\bar{y}}
\def\bz{\bar{z}}
\def\bH{\bar{H}}
\def\sl{$SL(2,R)$}
\def\slc{$SL(2,\mathbb{R})/U(1)$}
\def\nn{\nonumber}
\def\G{\Gamma}
\def\a{\alpha}
\def\at{\tilde{\alpha}}
\def\u{\Upsilon}

\def\Li{\textrm{Li}_2}
\def\h3{$\textrm{H}_3^+$}

\newcommand{\lm}{\lambda}
\def\d{{\partial}}
\def\vev#1{\left\langle #1 \right\rangle}
\def\cn{{\cal N}}
\def\co{{\cal O}}
\def\IC{{\mathbb C}}
\def\IR{{\mathbb R}}
\def\IZ{{\mathbb Z}}
\def\RP{{\bf RP}}
\def\CP{{\bf CP}}
\def\Poincare{{Poincar\'e }}
\def\tr{{\rm tr}}
\def\tp{{\tilde \Phi}}

\def\TL{\hfil$\displaystyle{##}$}
\def\TR{$\displaystyle{{}##}$\hfil}
\def\TC{\hfil$\displaystyle{##}$\hfil}
\def\TT{\hbox{##}}
\def\HLINE{\noalign{\vskip1\jot}\hline\noalign{\vskip1\jot}} 
\def\seqalign#1#2{\vcenter{\openup1\jot
  \halign{\strut #1\cr #2 \cr}}}
\def\lbldef#1#2{\expandafter\gdef\csname #1\endcsname {#2}}
\def\eqn#1#2{\lbldef{#1}{(\ref{#1})}%
\begin{equation} #2 \label{#1} \end{equation}}
\def\eqalign#1{\vcenter{\openup1\jot
    \halign{\strut\span\TL & \span\TR\cr #1 \cr
   }}}
\def\eno#1{(\ref{#1})}
\def\href#1#2{#2}

\def\ct{\cos t}
\def\st{\sin t}
\def\chr{ \cosh \rho}
\def\shr{ \sinh \rho}
\def\sp{ \sin \phi}
\def\cp { \cos \phi}

\def \perm {\pm ~{\rm perm.}}
\newcommand{\nca}{\newcommand}
\nca{\beq}{\begin{equation}} \nca{\eeq}{\end{equation}}
\nca{\beqa}{\begin{eqnarray}} \nca{\eeqa}{\end{eqnarray}}
\nca{\lsim}{\begin{array}{c}\,\sim\vspace{-21pt}\\< \end{array}}
\nca{\gsim}{\begin{array}{c}\sim\vspace{-21pt}\\> \end{array}}
\def\Dslash{\not{\hbox{\kern-3pt $D$}}}



\def\ads{{\it AdS}}
\def\adsp{{\it AdS}$_{p+2}$}
\def\cft{{\it CFT}}

\def\be{\begin{eqnarray}}
\def\ee{\end{eqnarray}}
\def\SS{\scriptscriptstyle}
\newcommand{\cfa}[4]{\!{\textstyle \left[ {{#1}\atop {#2}} {{#3}\atop
    {#4}} \right] }}

\newcommand{\cH}{{\cal H}}
\newcommand{\cG}{{\cal G}}
\newcommand{\cN}{{\cal N}}
\newcommand{\cO}{{\cal O}}
\newcommand{\cA}{{\cal A}}
\newcommand{\cT}{{\cal T}}
\newcommand{\cF}{{\cal F}}
\newcommand{\cC}{{\cal C}}
\newcommand{\cR}{{\cal R}}
\newcommand{\cW}{{\cal W}}
\def\cD{{\cal D}}
\def\cCo{{\cal C}}


\section{Introduction}

Black Holes continue to be a fascinating subject for study in string theory.
A central question is to understand the microstates of these black holes and compare
 their counting  with  the Bekenstein-Hawking entropy. This was first done  for big black holes in
$5$ dimensions in the classic work of Strominger and Vafa \cite{Strominger:1996sh}.
There have been several
important subsequent developments, see for example the reviews,
 \cite{Maldacena:1996ky},  \cite{hep-th/9710046}, \cite{Aharony:1999ti},
 \cite{hep-th/0203048}, and references therein.
Subleading corrections
have been analysed more recently,(see \cite{hep-th/9812082}, \cite{hep-th/9904005},
\cite{hep-th/9910179}, \cite{hep-th/9906094}, \cite{hep-th/0009234}),
\cite{hep-th/0412287},
and related to the topological  string partition function in
\cite{Ooguri:2004zv}. For small black holes the pioneering work was done by Sen,
(See \cite{hep-th/9504147} and \cite{hep-th/0505122})
and developed further with precise agreement being found between the microstate counting and
the Bekenstein-Hawking-Wald entropy in \cite{Sen:1997is}, \cite{Dabholkar:2004yr},
\cite{Dabholkar:2004dq}, \cite{Dabholkar:2005dt}.

The microscopic descriptions that have been developed  so far are usually in terms of a $1+1$
dim. Conformal Field Theory (CFT). Furthermore the microscopic counting has been done most reliably in
the thermodynamic limit of the CFT, see e.g., \cite{Strominger:1996sh},
\cite{Callan:1996dv},  \cite{Maldacena:1997de}, and the reviews, \cite{Maldacena:1996ky},
\cite{hep-th/9710046}, \cite{Aharony:1999ti},  \cite{hep-th/0203048}, and references therein;
some papers which discus the microscopic counting for non-supersymmetric black holes
are\footnote{For recent developments on rotating black holes, see, \cite{Emparan:2006it},
\cite{Emparan:2007en},\cite{Horowitz:2007xq}.}, \cite{Horowitz:1996ac}, \cite{Dabholkar:1997rk}, \cite{Dabholkar:2006tb}.
 In terms of the energy, $L_0$, and  central charge of the
CFT, $C$, the condition for the thermodynamic limit to be valid  takes the form,
\beq
\label{condcardy}
L_0\gg C.
\eeq
For a supersymmetric or non-supersymmetric extremal black hole, $L_0$ and $C$ are determined
by the charges carried by the black hole.
The entropy in this limit is  given by the well known  Cardy formula,
\beq
\label{entcardy}
S=2 \pi \sqrt{C L_0\over 6}.
\eeq
In the discussion below, we will often refer to the thermodynamic limit as the Cardy limit.
We see from eq.(\ref{entcardy}) that  in this limit a knowledge of the central charge
and the energy, $L_0$, is sufficient
to determine the entropy. Moreover, the central charge is  a robust quantity which can
often be determined quite easily by anomaly considerations. This makes
it easy to carry out a microscopic calculation of the entropy, \cite{Sfetsos:1997xs},
\cite{hep-th/9712251}, \cite{Skenderis:1999bs}.

In addition, when the condition, eq.(\ref{condcardy}) is valid subleading corrections to the entropy can also  often be easily calculated. These continue to have the form, eq.(\ref{entcardy}). The subleading corrections arise due to corrections to the central charge, $C$ and can be determined
by anomaly considerations\cite{Kraus:2005vz}, \cite{Kraus:2005zm}, \cite{Kraus:2006nb}, \cite{Kraus:2006wn}.

Since so much can be understood in the Cardy limit, it is natural to ask whether any charge configuration can be put in the Cardy limit using the duality symmetries of string theory. This is the main question we will explore in this paper. Our focus is  on big black holes.
These carry  large charges, $Q\gg 1$,
and have a horizon radius which is large compared to the Planck and string scales, so that
their horizon
geometry is  well described by the supergravity approximation.
We are   interested in both supersymmetric and non-supersymmetric extremal black holes
 of this type.

We will focus on black holes obtained in    Type IIA string theory compactified on $K3 \times T^2$,
with duality group, $O(6,22,{\mathbb Z}) \times SL(2,{\mathbb Z})$.
For a configuration with $D0-D4$ brane charges we identify some necessary conditions which
must be met. Generically, it turns out
that these conditions  cannot be met, leading to the conclusion that a generic set of charges
cannot be taken to the Cardy limit.  These results are valid for both supersymmetric and
non-supersymmetric extremal black holes.
We find that  the required non-genericity, to be able to take a set of charges to the Cardy limit, is
interestingly different in the two cases.  In the non-supersymmetric case, unlike the supersymmetric one,
 a ``near-by'' charge configuration can always be found which can be brought to the Cardy
limit.  The fractional shift in the charge required to go to the near-by configuration, satisfies the
condition,
\beq
\label{introconda}
{\Delta Q \over Q} \sim {1\over \sqrt{Q}},
\eeq
and is small for large charge.
 Similar results are also shown to hold in the $D0-D6$ system, which is non-supersymmetric.
In this case one can never take the charges to the Cardy limit, but again,   a small alteration
in the charges brings us to a $D0-D2-D4-D6$ system which can be taken to the Cardy limit.
 Our results can be extended to some more general charges  in a straightforward way.
We also expect similar results to hold in other compactifications, for example of Type IIA on
$T^6$,
and  Heterotic theory on $K3\times T^2$.

It is important to emphasise that the  results mentioned above arise because
the duality group is  discrete.
  If instead of $O(6,22,{\mathbb Z})\times SL(2,\mathbb Z)$ we  consider the continuous group,
$O(6,20, \mathbb R)\times SL(2,\mathbb R)$,  then
it is well known that it  is always possible to bring a configuration
with large charges \footnote{By large charge we mean that both $Q\gg 1$ and $I\gg 1$.}
to the Cardy limit.  The continuous group has only one invariant,
 $I=\vec{Q}_e^2 \vec{Q}_m^2-(\vec{Q}_e\cdot \vec{Q}_m)^2$, where, $\vec{Q}_e, \vec{Q}_m$ are
the $28$ dimensional electric and magnetic charge vectors.
Thus any  set of charges, $(\vec{Q}_e, \vec{Q}_m)$, can always be transformed to  one
 in the Cardy limit, with the same value
of this invariant.
 The discrete group is smaller
and there are additional discrete invariants that characterise its representations.
It should be possible to understand the obstruction to bringing a general set of charges to the
 Cardy limit in terms of these additional  invariants and also understand the
required non-genericity in  terms of these invariants.
 We leave this more complete analysis for the future.

If the charges lie in the Cardy limit, the  black hole  admits a description as a BTZ black hole in $AdS_3$,  in some region of moduli space. It can therefore be regarded as a state in a $1+1$ dim. CFT one and its entropy is given by the Cardy formula, eq.(\ref{entcardy}). Our result, that a generic non-supersymmetric state, after a small shift in charges, can be brought to the Cardy limit, thus tells us that at least in some region of moduli space the entropy of the corresponding black hole can be understood microscopically.

This is a promising start but one would like to do better. In fact the long-term goal behind this work is  to try and get an understanding of entropy for four- dimensional extremal non-supersymmetric black holes. The near horizon geometry of these black holes  is $AdS_2\times S^2$. In some cases an internal circle  combines with the $AdS_2$ component
giving rise to a locally $AdS_3$ space, but even in these cases generically the charges  do not lie in the Cardy limit.
What our result shows is that at least in some region of moduli space,  the entropy of such a black hole can be understood microscopically. In this region of moduli space the geometry is that of a BTZ black hole in $AdS_3$ space. 
We discuss in the conclusions how an argument might be developed with this starting point, leading to a microscopic derivation of the entropy in other regions of moduli space where the black hole is four dimensional. Such an argument should also be applicable to rotating black holes, including the extreme Kerr black hole in four dimensions.

One comment is worth making at this stage \footnote{We thank
S. Mathur and A. Strominger for emphasising this point to us.}. Sometimes the condition eq.(\ref{condcardy}) \  is not necessary and a much weaker condition suffices.
This happens for example in the D1-D5-P system when the CFT is at the orbifold point. At this point in the moduli space the twisted sectors can be thought of as multiply wound strings. In the singly wound sector the relevant condition is given by eq.(\ref{condcardy}). In contrast in the maximally
wound sector the effective central charge is order one and energy is given by replacing  $L_0$ by,
\beq
\label{replacecl}
L_0 \rightarrow L_0 Q_1Q_5,
\eeq
where $Q_1,Q_5$ are the $D1, D5$ brane charges.
Thus the condition, eq.(\ref{condcardy}), is automatically met for large charges in the maximally wound sector.

Away from the orbifold point though the different twisted sectors mix. The only condition which  can now  guarantee the validity of the Cardy formula is  eq.(\ref{condcardy}), which ensures that the system is in the thermodynamic limit. It is well known that the CFT dual to the Black hole is not at the orbifold point. Thus a microscopic calculation of the entropy using the Cardy formula would require this condition  to be valid. In the supersymmetric case, where one is calculating an index, one can still justify working at the orbifold point, where the dominant contribution comes from the maximally wound sector, 
 and hence one would not need to impose the condition, eq.(\ref{condcardy}).
However, for non-supersymmetric black holes, which are the ones of primary interest in   this paper, the entropy can change as one moves in moduli space.  A legitimate microscopic
calculation in this case would have to be done away from the orbifold point and would require the condition, eq.(\ref{condcardy}),  to hold for the Cardy formula to be valid.

It should be mentioned that the mass gap for excitations above the BTZ black hole can be calculated in the gravity side and is well known to  go like,
\beq
\label{massgap}
E_{gap}\sim 1/(L C),
\eeq
where $L$ is the length of the circle on which the CFT lives. This shows that an effective picture in terms of one multiply wrapped long string  must continue to hold even away from the orbifold point. However a first principles argument of why this happens is still missing especially in the non-supersymmetric case.
In the absence of such an argument it is appropriate  to require,  at least in  a first principles calculation of the microscopic entropy, that for the Cardy formula to be valid the condition, eq.(\ref{condcardy}), holds. This paper explores how restrictive this condition is, once the duality symmetries of string theory are taken into account.

The paper is organised as follows. We start with some background in \S2.
In \S3, we discuss the $D0-D4$ system, and in \S4, the $D0-D6$ system.
In \S5, we prove that for the lift in M-theory to give a locally $AdS_3$ space
the $D6$-brane charge must vanish.  We end with some conclusions in \S6. The appendices, A-D,
contain supporting results and discussion.

\section{Background}

The compactification of Type IIA theory on $K3\times T^2$ preserves $16$ supersymmetries.
It is dual to Heterotic theory on $T^6$\cite{hep-th/9802051}.
The resulting four dimensional theory has $28$ gauge fields. In the IIA description these arise
as follows. One gauge field comes from the RR 1-form gauge potential, $C_1$;
$23$ gauge fields from the KK reduction of the RR 3-form gauge potential, $C_3$,
on the 22 non-trivial 2-cycles
of $K3$ and on the $T^2$; and $4$ gauge fields from the KK reduction of the metric and the 2-form
NS field, $B_2$ on the  1-cycles of the $T^2$.
The duality group is $O(6,22, {\mathbb Z}) \times SL(2,{\mathbb Z})$.  $O(6,22,{\mathbb Z})$ is the T-duality group
of the Heterotic theory, and $SL(2,{\mathbb Z})$ is the S-duality symmetry of the
4 dimensional Heterotic theory.

A general state carries electric and magnetic charges  with respect to these gauge fields.
The electric charges, $\vec{Q}_e$, and the magnetic charges, $\vec{Q}_m$,
take values in a lattice, $\Gamma^{6,22}$, which is even, self-dual and of signature, $(6,22)$.
The lattice is invariant under the group, $O(6,22,\mathbb Z)$. The electric and magnetic charges,
$\vec{Q}_e, \vec{Q}_m$, transform
 as  vectors of $O(6,22,\mathbb Z)$. And together, $(\vec{Q}_e, \vec{Q}_m)$, transform as a doublet
of $SL(2,\mathbb Z)$.
In a particular basis, $\{e_i\}$ of $\Gamma^{6,22}$, the matrix of inner products,
\beq
\label{inpro}
\eta_{ij}\equiv (e_i, e_j),
\eeq
takes the form,
\beq
\label{metric}
\eta= {\cal H} \oplus {\cal H} \oplus {\cal H} \oplus {\cal H} \oplus {\cal E}_8
 \oplus {\cal E}_8 \oplus {\cal H} \oplus {\cal H}.
\eeq
Here $\cal{H}$, is given by,
\beq
\label{defH}
{\cal H}=\pmatrix{0 & 1 \cr 1 & 0 \cr },
\eeq
and ${\cal E}_8$ is the Cartan matrix of $E_8$.

In this basis, the electric charge vector has  components,
\beq
\label{defqe}
\vec{Q}_e=(q_0, -p^1, q_i, n_1, NS_1, n_2, NS_2).
\eeq
Here, $q_0$ is the $D0$-brane charge; $p^1$ is the charge due to $D4$-branes wrapping $K3$;
$q_i, i=2, \cdots 23$ are
the charges  due to D2-branes wrapping the 22 2-cycles of $K3$  which we denote as
$C_i$;
$n_1, n_2$ are the momenta along the two 1-cycles of $T^2$ and $NS_1, NS_2$ are  the charges due to
$NS_5$ branes wrapping $K3\times S^1$ where $S^1$ is one of the two 1-cycles of $T^2$.

And the magnetic charge vector has components,
\beq
\label{defqm}
\vec{Q}_m=(q_1, p^0, p^i, w_1, KK_1, w_2, KK_2).
\eeq
Here, $q_1$ is the charge due to $D2$-branes wrapping $T^2$; $p^0$ is the $D6$-brane charge; $p^i, i=2
\cdots 23$, are  the charges
due to $D4$-branes wrapping the cycle ${\tilde C_i} \times T^2$, where ${\tilde C_i}$ is the 2-cycle
on K3 dual to $C_i$; $w_1,w_2$ are charges due to the winding modes of the fundamental string
along the two 1-cycles
of $T^2$; and $KK_1,KK_2$ are the KK-monopole charges  that arise  along the two 1-cycles of the $T^2$.

Three bilinears in the charges can be defined,
\beqa
\label{biforms}
\vec{Q}_e^2 & \equiv &  (\vec{Q}_e,\vec{Q}_e) \cr
\vec{Q}_m^2  & \equiv &  (\vec{Q}_m,\vec{Q}_m) \cr
\vec{Q}_e \cdot \vec{Q}_m & \equiv &  (\vec{Q}_e,\vec{Q}_m).
\eeqa
These are invariant under  $O(6,22, \mathbb Z)$.

An invariant under the full duality group is,
\beq
\label{definv}
I=(\vec{Q}_e)^2(\vec{Q}_m)^2-(\vec{Q}_e\cdot \vec{Q}_m)^2.
\eeq
It is quartic in the charges.
For a big supersymmetric black hole,  $I$ is positive, and the entropy of the black hole\cite{hep-th/0507014} is,
\beq
\label{entsusy}
S=\pi{\sqrt{\vec{Q}_e^2\vec{Q}_m^2-(\vec{Q}_e\cdot \vec{Q}_m)^2}}.
\eeq
In contrast, for a big non-supersymmetric extremal black hole, $I$ is negative and the entropy is,
\beq
\label{entnonsusy}
S=\pi \sqrt{(\vec{Q}_e\cdot \vec{Q}_m)^2 - \vec{Q}_e^2\vec{Q}_m^2}.
\eeq

We now turn to discussing  the Cardy limit.
Consider  a Black hole carrying $D0-D4$ brane charge. In our notation
the non-zero charges are, $q_0, p^1, p^i,i=2, \cdots 23$.  This solution can be lifted to
M-theory, and the near horizon geometry in M-theory is given by a BTZ black hole in
$AdS_3\times S^2$. The $AdS_3$ space-time admits a dual description in terms of a $1+1$ dim. CFT
living on its boundary. The central charge, $C$, of the CFT can be calculated from the bulk,
it is determined by the curvature of  the $AdS_3$ spacetime. For large charges we get,
\beq
\label{cd0d4}
C=3|p^1 d_{ij} p^ip^j|,
\eeq
where $d_{ij}$ is the  matrix $\eta_{ij}$, eq.(\ref{inpro}), restricted to the
$22$ dimensional subspace of charges given by  $D4$-branes wrapping two-cycles of $K3$ and
 $T^2$. This    corresponds to the second, third and fourth factor of
$\cal{H}$ and the two ${\cal E}_8$'s in eq.(\ref{metric}).

 The BTZ black hole is a quotient of $AdS_3$ obtained by identifying points separated
by a space-like direction. The symmetry of $AdS_3$ is $SO(2,2)$; this is broken by the
identification of points in the BTZ black hole  to $SO(2,1) \times U(1)$.
The size of the circle obtained by this identification, $L$,
 is given in terms of the radius of $AdS_3$, $R_{AdS}$, by
\beq
\label{lena}
{L \over R_{AdS}}\sim {|q_0| \over C},
\eeq
where $q_0$ is the zero-brane charge carried by the Black hole.

In the Cardy limit the condition,
\beq
\label{cardycondback}
|q_0|\gg C,
\eeq
is satisfied.
From eq.(\ref{lena}) we see that this leads to the condition,   ${L \over R_{AdS}} \gg 1$. From,
eq.(\ref{cd0d4}) we see for this limit to be valid, the condition,
\beq
\label{cardyd0d4}
|q_0| \gg |p^1 d_{ij} p^ip^j|,
\eeq
must hold.
Since, ${L \over R_{AdS}} \gg 1$, in the Cardy limit, the distance between points  which are identified
in the BTZ background is much bigger than $R_{AdS}$.
 As a result,  the effect of  the reduced symmetry in the BTZ background,
 due to taking the quotient,   can be neglected in the Cardy limit.
The partition function in the bulk can then  be calculated using the full symmetries
of $AdS_3$. The resulting answer  is the well known Cardy formula,
\beq
\label{cardyent}
S=2\pi \sqrt{C |q_0| \over 6}.
\eeq
The Cardy limit corresponds to the thermodynamic limit of the microscopic $1+1$ dim. CFT.
In this limit the dimensionless temperature $T$ of the CFT satisfies the condition,
\beq
\label{ctemp}
T \gg 1.
\eeq
Away from the Cardy limit the breaking of $SO(2,2)$ to $SO(2,1)$ becomes important and there is
no way to calculate the partition function or entropy without knowing more details of the bulk, or
the dual boundary conformal field theory.

So far we have considered a system with $D0-D4$ brane charge. What about including other charges?
 If a $D6$-brane charge is also present,
 we show in \S5, that on lifting to M-theory one does not get an $AdS_3$
space-time.
 All other charges are allowed by the requirement that the M-theory lift gives
an $AdS_3$ spacetime in the near-horizon limit.
So  a general configuration which admits
an $AdS_3$ lift can also include $D2$-brane charges, and non-zero values
for $n_1,n_2, w_1,w_2, NS_1,NS_2, KK_1, KK_2$, besides having $D0-D4$ brane charges.
The resulting central charge of the $1+1$ dim. CFT after lifting to M-theory is \footnote{
The central charge is determined by all the branes which are extended strings in the $AdS_3$.
One can see from eq.(\ref{defqe}), eq.(\ref{defqm}), that
 this formula gives a dependence on all of them. Localised excitations, like momentum modes or
 wrapped 2-branes,  correspond to states and do not change the central charge.}
\beq
\label{cchargegen}
C=3 |p^1 \vec{Q}_m^2|.
\eeq

In  the more general case, the condition for the Cardy limit is,
\beq
\label{cardygen}
|\hat{q_0}| \gg C.
\eeq
Where, $|\hat{q_0}|$ is,
\beq
\label{defq0}
|\hat{q_0}| ={|\vec{Q}_e^2 \vec{Q}_m^2 -(\vec{Q}_e \cdot \vec{Q}_m)^2| \over 2 |p^1 \vec{Q}_m^2|}.
\eeq
Using eq.(\ref{definv}), eq.(\ref{cchargegen}) and eq.(\ref{defq0}) this can be written in the form,
\beq
\label{two}
I\gg 6  (p^1)^2 (\vec{Q}_m^2)^2.
\eeq
To summarise, for a charge configuration  to be in the Cardy limit,  two conditions
must  hold.
First the $D6$-brane charge, $p^0$, must vanish. Second, eq.(\ref{cardygen}) or equivalently,
eq.(\ref{two}),  must be valid.
We refer to these two conditions as the Cardy conditions below.

Before proceeding let us note that we are neglecting $1/Q$ corrections in
the formula for the central charge, eq.(\ref{cchargegen}).
 For these to be small, the
 BTZ black hole should be  a state in a weakly coupled $AdS_3$ background.
 The Radius of the
$AdS_3$ space, $R_{AdS}$, in units of the three dimensional Planck scale, $l_{Pl}^{(3)}$,
is given by,
\beq
\label{clargeconda}
{R_{AdS} \over l_{Pl}^{(3)}}\sim C.
\eeq
For the  BTZ black hole to be a state in a weakly coupled $AdS_3$ spacetime,
${R_{AdS}\over l_{Pl}^{(3)}}\gg 1$, yielding the condition\footnote{The stronger conditions
are, ${R_{AdS}\over l_{11}} \gg 1$, ${R_{S^2}\over l_{11}} \gg 1$, and
${V_6\over l_{11}^6}\gg 1$,
where $R_{S^2}, V_6$ are the Radius of the $S^2$ and volume of the internal space respectively.
From these, and the relation, $l_{Pl}^{(3)}={l_{11}^9 \over R_{S^2}^2 V_6}$, the condition,
${R_{AdS}\over l_{Pl}^{(3)}}\gg 1$, follows.},
\beq
\label{clargecondb}
C \gg 1.
\eeq

The conditions on the charges for the Cardy limit are not duality invariant. This raises the
question, when can a charge configuration be brought to the Cardy limit after a duality transformation?
This is the central question we address in this paper.
In \S3 we first address this question for the case where the starting configuration,
has $D0-D4$ brane charges. Our analysis includes both the supersymmetric and non-supersymmetric cases.
Following this in \S4, we address this question when the starting configuration carries
$D0-D6$ brane charges.

There is one potentially confusing point that we would like to  address before going further.
In asking whether a system of charges can be brought to the Cardy limit,
we are really asking whether  any of the   internal circles  of the compactification can combine
with the $AdS_2$ component of the near horizon  geometry and give rise to  a
 three-dimensional BTZ black hole and whether this black hole has charges which lie in the Cardy limit.
There are six  internal circles for example in the Heterotic description, corresponding to the
$6$ Hyperbolic lattices, $\cal{H}$ in eq.(\ref{metric}), and we  allow for the internal
circle to be any one of them.   Our results, mentioned in the introduction, which say that generically
this is not possible, mean that for generic charges
there is no internal circle  which can combine in this manner, yielding
the Cardy limit.

There are two ways to carry out the analysis. We can keep the charges fixed and ask
whether a suitable circle can be found.  This corresponds to a passive transformation,
under which the charges are kept fixed but the basis in the charge lattice,
 with respect to which the components were written in eq.(\ref{defqe}),
eq.(\ref{defqm}), is changed. Alternatively, we can keep the basis fixed and change the charges,
and ask whether the transformed charges meet  the required conditions. This corresponds
to an active transformation. We will adopt this latter active of point of view in the paper.
In this point of view the internal circle which combines and  potentially gives rise to a BTZ black
hole is  kept fixed and in our conventions is  the M-theory circle in the IIA description.

\section{The $D0-D4$ System}

In this section we analyse the $D0-D4$ system. Subsection 3.1  discusses the supersymmetric case,
and subsections 3.2, 3.3,  discuss the non-supersymmetric case. In both cases we  find that
a generic set of charges cannot be brought to the Cardy limit. Subsection 3.4,  discuss what happens
if starting with generic charges we now allow the charges to vary. We  find that in the non-supersymmetric
case a near-by charge configuration can always be found which can be brought to the Cardy limit.
Additional relevant material  is in appendices $A$ and $B$.

Our starting configuration for the $D0-D4$ case has non-zero values for
$q^0, p^1, p^i$, in the notation of eq(\ref{defqe}), eq.(\ref{defqm}), and
all other charge are vanishing.
It is easy to see from eq.(\ref{biforms}) that
\beq
\label{eminv}
\vec{Q}_e\cdot \vec{Q}_m=0,
\eeq
 in this case.

In our analysis  we are interested in the case of large charges, $|q_0|, |p^1|, |p^i| \gg 1$.
The Cardy condition for the starting configuration takes the form,
eq.(\ref{cardyd0d4}). We see that for a generic set of initial charges this condition will not be met.
Generically all charges will be roughly comparable, $|q_0|\sim |p^1|\sim |p^i| \sim Q \gg 1$
Now the LHS of eq.(\ref{cardyd0d4}) is  linear in $Q$ while the RHS is cubic in $Q$, so
generically, for $Q \gg 1$,
 the inequality, eq.(\ref{cardyd0d4}),
will not be met.

Below we formulate a set of necessary condition which must be met, for the final configuration to be in
the Cardy limit. For generic initial charges, we find that these conditions are not met.
And so we learn that generically a system with $D0-D4$ charge cannot be brought to the Cardy limit.
In some special, non-generic cases, these necessary conditions are  met.
We construct some examples   of this type and explicitly   find a  duality transformation
bringing them to the Cardy limit \footnote{Of course a trivial way in which this could happen is if the
initial configuration, while being non-generic, is itself in the Cardy limit, and meets condition,
eq.(\ref{cardyd0d4}). In the example we construct,  the initial charges while being rather
special are not   in the Cardy limit. We find explicitly the duality transformation
 bringing them to this limit.}.

Let us denote the final configuration which is obtained after carrying out a duality transformation
on the initial $D0-D4$ charges by $(\vec{Q}_e', \vec{Q}_m')$.
As was pointed out above, the $D6$-brane charge, $p^{0'}$, in the final configuration must vanish for this
to happen, and eq.(\ref{two}) must be met.

We can restate eq.(\ref{two}) in the slightly weaker form as,
\beq
\label{conda}
|I| \gg (p^{1'} (\vec{Q}'_m)^2)^2.
\eeq
This gives rise to the condition,
\beq
\label{condaalt}
\left|{(\vec{Q}_m')^2 \over \sqrt{|I|}}\right|\ll {1\over |p^{1'}|}.
\eeq
Since $|p^{1'}| >1$ eq.(\ref{condaalt}) leads to the condition,
\beq
\label{condb}
\left|{(\vec{Q}_m')^2 \over \sqrt{|I|}}\right| \ll 1.
\eeq

The final configuration, $(\vec{Q}_e', \vec{Q}_m')$ is obtained from the initial one, by the action of
a combined $SL(2,\mathbb Z)$ transformation and an $O(6,22,\mathbb Z)$ transformation. Denote the element of
$SL(2,\mathbb Z)$ by
\beq
\label{elsl2z}
A=\pmatrix{ a & b \cr c & d \cr}.
\eeq
By definition, $a,b,c,d, \in {\mathbb Z}$ and $ad-bc=1$.
The $SL(2,\mathbb Z)$ transformation acts on the charges as follows,
\beqa
\label{actcharges}
\vec{Q}_e & \rightarrow &  a \vec{Q}_e+b \vec{Q}_m \cr
\vec{Q}_m & \rightarrow & c\vec{Q}_e+d\vec{Q}_m.
\eeqa

The $O(6,22)$ transformation does not change the value of the bilinears, eq.(\ref{biforms}),
also  the initial charges satisfy the condition,  $\vec{Q}_e \cdot \vec{Q}_m=0$. This leads to,
\beq
\label{rela}
(\vec{Q}_m')^2  = c^2 \vec{Q}_e^2 + d^2 \vec{Q}_m^2.
\eeq
Using eq.(\ref{condb}), now gives,
\beq
\label{condc}
\left|c^2  {\vec{Q}_e^2 \over\sqrt{|I|}} + d^2 {\vec{Q}_m^2 \over \sqrt{|I|}}\right| \ll 1.
\eeq
This condition will play an important role in the discussion below.

\subsection{The Supersymmetric Case}

Since eq.(\ref{eminv}) is true for the $D0-D4$ system,
it follows from eq.(\ref{definv}) that the duality invariant, $I$, is,
\beq
\label{valI}
I=\vec{Q}_e^2 \vec{Q}_m^2.
\eeq
For a supersymmetric system, $I>0$, so we see that
 $\vec{Q}_e^2, \vec{Q}_m^2$ have the same sign.
From, eq.(\ref{rela}) it follows  that $(\vec{Q}_m')^2$ must also have the same
 sign as $\vec{Q}_e^2, \vec{Q}_m^2$.

Thus eq.(\ref{condb}) takes the form,
\beq
\label{condctwo}
c^2 {|\vec{Q}_e^2|\over \sqrt{I}} + d^2 {|\vec{Q}_m^2| \over \sqrt{I}} \ll 1.
\eeq

Now by doing an $SL(2,\mathbb Z)$ transformation if necessary we can always take
the initial charges to satisfy the condition,
\beq
\label{rata}
\left|{\vec{Q}_e^2 \over \vec{Q}_m^2}\right| \ge 1.
\eeq
(Either this condition is already met or we do the $SL(2,\mathbb Z)$ transformation
$(\vec{Q}_e, \vec{Q}_m) \rightarrow (-\vec{Q}_m, \vec{Q}_e)$ after which it is true).

Using the expression for $I$ in eq.(\ref{valI}), eq. (\ref{rata}) leads to,
\beq
\label{ratb}
{|\vec{Q}_e^2| \over \sqrt{I}}  \ge 1.
\eeq
Now since $c,d$ are integers, we see that the only way, eq.(\ref{condctwo}) can be met is if, $c=0$.
The resulting $SL(2,\mathbb Z)$ matrix must then take the form,
\beq
\label{slfirst}
A=\pmatrix{1 & b \cr  0 & 1 \cr}.
\eeq
From  eq.(\ref{rela})  it now follows that,
\beq
\label{mpc}
(\vec{Q}_m')^2=\vec{Q}_m^2.
\eeq

The condition, eq.(\ref{condb}), using eq.(\ref{valI}), eq.(\ref{mpc})  then leads to,
\beq
\label{condfe}
|\vec{Q}_e^2| \gg |\vec{Q}_m^2|
\eeq

A few points are now worth making. Eq.(\ref{condfe})
is a necessary condition  on the initial charges $(\vec{Q}_e,\vec{Q}_m)$  which
must be met, to be able to go to the Cardy limit.  It is easy to see that
this  condition will not be met generically.
If all the initial charges, $q_0, p^1, p^i$ are of the same order, $Q \gg 1$,
 then, $\vec{Q}_m^2=2 d_{ij}p^ip^j$
and $\vec{Q}_e^2=-2 q_0p^1$  are both quadratic in $Q$ and will generically be roughly comparable, so that
eq.(\ref{condfe}) is  not met. On the other hand this condition
is somewhat less non-generic than the condition required for the initial configuration to be in the Cardy limit, since both sides of the inequality scale like $Q^2$ in eq.(\ref{condfe}), while
in  eq.(\ref{cardyd0d4}) the rhs scales relative to the lhs by a factor of $Q^2$.
Thus one can  find initial charges which are not in the Cardy limit, but
which meet the  condition eq.(\ref{condfe}). We will present some explicit examples
below and show that they can be sometimes
brought to  the Cardy limit by duality transformations.

Before doing so let us comment that the  eq.(\ref{condfe})
 can in fact be somewhat tightened.
Let ${\rm gcd}(\vec{Q}_e)$ stand for the greatest common divisor  of all the integer charges in $\vec{Q}_e$.
Then the stronger form of this condition is,
\beq
\label{sforma}
|\vec{Q}_e^2| \gg ({\rm gcd}\vec{Q}_e)^2 |\vec{Q}_m^2|
\eeq
In Appendix A, we discuss how eq.(\ref{sforma}) can be derived.

In the  example we present next, the starting configuration is not in the Cardy limit,
but condition, eq.(\ref{sforma}) is met. We will present the explicit duality transformation
that brings this configuration to the Cardy limit.

\subsubsection{An Explicit Example}
We start with  the charges,
\beq
\label{exa}
\vec{Q}_e=(-p^1+1,-p^1, 0,0, 0,0,0, \cdots, 0)
\eeq
\beq
\label{exb}
\vec{Q}_m=(0,0,p^2,p^2, 0 , 0,0,  \cdots , 0)
\eeq
with,
\beq
\label{exconda}
(p^1)^2 \gg 3 (p^2)^2 \gg 1.
\eeq
The quadratic bilinears, eq.(\ref{biforms}), take the values,
\beqa
\label{biex}
\vec{Q}_e^2 & = & 2(p^1-1)p^1 \cr
\vec{Q}_m^2 & = & 2 (p^2)^2  \cr
\vec{Q}_e\cdot \vec{Q}_m & = & 0
\eeqa

The invariant, $I$, eq.(\ref{definv}), takes the value,
\beq
\label{valiex}
I=4p^1(p^1-1) (p^2)^2
\eeq
Note that this starting configuration is not in the Cardy limit as  these charges do
 not satisfy  the condition, eq.(\ref{two}). But the starting configuration does satisfy
 eq.(\ref{sforma}) since,
${\rm gcd}(\vec{Q}_e)= {\rm gcd}(p,p-1)=1$, and eq.(\ref{exconda}) holds.

Now we carry out the transformation, $B\in O(3,3,\mathbb Z) \subset O(6,22,\mathbb Z)$, given by,
\beq
\label{bmatex}
B= \left(\matrix{ 0 & 1 & 0 & 0 & 0 & 0 \cr
                  1 & -1 & 1 & -1 & 2 & 1 \cr
                  0 & -1 & 0 & 1 & 0 & 0  \cr
                  0 & 0 & 1 & 0 & 1 & 0   \cr
                  0 & -1 & 0 & 0 & 1 & 0  \cr
                  0 & -1 & 0 & -1 & 0 & 1 } \right) ~.
\eeq
 $B$ acts non-trivially on the $6$ dimensional sublattice of $\Gamma^{6,22}$,
with an inner product given by
 first three $\cal{H} \oplus \cal{H} \oplus \cal{H}$ factors in  eq.(\ref{inpro}),
 and acts trivially on the rest of the lattice.
The resulting charges are given by,
\beq
\label{exfe}
\vec{Q'_e}=(-p^1,1,p^1,0,p^1,p^1, 0,\cdots, 0)
\eeq
\beq
\label{exfm}
\vec{Q'_m}=(0,0,p^2,p^2,0,-p^2, 0, \cdots 0)
\eeq
Since the second entry in $\vec{Q'_m}$ vanishes, there is no $D6$-brane charge.
From, eq.(\ref{exfe}) we see that $p^{1'}=-1$.
Also,
\beq
\label{relqm}
\vec{Q'_m}^2=2 (p^2)^2.
\eeq

Now  the Cardy condition requires that,
\beq
\label{ccond}
I\gg 6 \left(p^{1'}\vec{Q'_m}^2\right)^2 ~.
\eeq
Using eq.(\ref{valiex}),  eq.(\ref{relqm}) and eq.(\ref{exconda}), we see that
 this condition is indeed met.

An example where all the final charges are much bigger than unity can be obtained
by scaling all the charges above, by $\lambda \gg 1$ and  taking
\beq
\label{newcond}
(p^1)^2 \gg 3 (\lambda)^2 (p^2)^2.
\eeq

\subsection{The Non-supersymmetric Case}
In the Non-supersymmetric $D0-D4$ system, $I$ also takes the form, eq.(\ref{valI}).
By doing an $SL(2,\mathbb Z)$ transformation if necessary we can assume,  without loss of generality
that
\beq
\label{ineq}
{|\vec{Q}_m^2| \over |\vec{Q}_e^2|} \le 1.
\eeq
For subsequent discussion it is useful to define the parameter, $\alpha$, as follows,
\beq
\label{defalpha}
\alpha={|\vec{Q}_m^2| \over \sqrt{|I|}}={\sqrt{|I|}\over|\vec{Q}_e^2|}=
\sqrt{|\vec{Q}_m^2| \over |\vec{Q}_e^2|}.
\eeq
where the last two equalities follows from eq.(\ref{valI}). We see from eq.(\ref{ineq}) that
\beq
\label{condaa}
\alpha \le 1.
\eeq

Since $I$ is negative, we learn from eq.(\ref{valI}) that $\vec{Q}_e^2, \vec{Q}_m^2$ must
 have opposite signs.
There are  then two possibilities,  either $\vec{Q'_m}^2$ has the same sign as $ \vec{Q}_e^2$,
or it has the opposite sign as $\vec{Q}_e^2$.
In both cases, eq.(\ref{condc}) takes the form,
\beq
\label{relnona}
0< \left|-d^2 \alpha + {c^2 \over \alpha}\right| \ll 1.
\eeq
The requirement  $|-d^2 \alpha + {c^2 \over \alpha}| >0$ arises from the condition
that ${\vec{Q'}_m}^2$ is non-vanishing, and this  in turn arises from the requirement that
the central charge, $C$, eq.(\ref{cchargegen}), does not vanish.

The analysis and conclusions are similar in the two cases.
 Below we give details for the
case when ${\vec{Q'}_m}^2$ and $\vec{Q}_e^2$ have the same sign and also state the
conclusions for the case when $\vec{Q}_m'^2$ and $\vec{Q}_e^2$ have the opposite sign.

In the  case when $\vec{Q}_m'^2, \vec{Q}_e^2,$ have the same sign,  eq.(\ref{relnona})
takes the form,
\beq
\label{relnonb}
0<-d^2\alpha +{c^2\over \alpha} \ll 1.
\eeq

It is interesting to compare this with the condition that arose in the susy case,
  eq.(\ref{condctwo}).
This  constraint  required the charges
to be non-generic and to  satisfy the condition,  eq.(\ref{condfe}), in the susy case.
In terms of $\alpha$,
defined in eq.(\ref{defalpha}), this condition takes the form,
\beq
\label{csusya}
\alpha^2 \ll 1.
\eeq
At first sight it might seem that the difference in  relative sign between the two terms
 makes eq.(\ref{relnonb}) easier to satisfy in the non-susy case. To explore this question
we will take, $\alpha<1$, but not much less than unity and ask whether such a set of
charges can be brought to the Cardy limit.
We will find that in fact eq.(\ref{relnonb}) cannot be met for generic initial charges.
Also, we will see that    the nature of the  non-genericity which allows eq.(\ref{relnonb}) to be met
 is interestingly different from
the susy case, and this has interesting consequences which we will discuss further in the next
subsection.

Conditions, eq.(\ref{condaa}) and eq.(\ref{relnona}), and the fact
that $c$ takes integer values,
 imply that $d$ cannot vanish.
We can then write eq.(\ref{relnonb}) as follows,
\beq
\label{connonc}
0<{d^2 \over \alpha}\left(-\alpha^2+{c^2\over d^2}\right) \ll 1.
\eeq
Since  $d^2\ge 1$ and $\alpha\le 1,$ this gives rise to a weaker condition,
\beq
\label{connond}
0 < \left(-\alpha +\left|{c\over d}\right|\right) \left(\alpha + \left|{c\over d}\right|\right)  \ll 1.
\eeq
Now if $\alpha$ is not very much less than unity, as we are assuming, then
$(\alpha + |{c\over d}|)$ cannot be very much less than unity. Thus the only way to meet the
condition, eq.(\ref{connond}), is for
\beq
\label{connone}
0< \left|{c\over d}\right| - \alpha \ll 1.
\eeq
In general we see from eq.(\ref{defalpha}) that $\alpha$
 is an irrational number and
$|{c\over d}|$ is a rational number. We know that any irrational number can be approximated
arbitrarily well by a rational number, therefore one can meet condition eq.(\ref{connone}) for
a general $\alpha$.

Let us however go back to the stronger condition, eq.(\ref{connonc}), we will see that
this cannot be met generically. We state the condition in
eq.(\ref{connonc}) as
follows:
\beq
\label{nsusyconnonc}
0<{d^2\over \alpha}\left(-\alpha^2+{c^2\over d^2}\right) < \delta,
\eeq
where,
 $\delta$ is a small number satisfying,
\beq
\label{nsusyg}
\delta \ll 1.
\eeq
Eq.(\ref{connone}) then takes the form,
\beq
\label{nsusyf}
0<\left|{c\over d}\right|-\alpha <\delta.
\eeq
As was mentioned above, since any irrational number can be approximated   arbitrarily well by a rational
number, $c,d$ can always be  found so that eq.(\ref{nsusyf}) is met.
However,  for  a  generic irrational number, $\alpha$,  the integers, $d,c$  that
satisfy eq.(\ref{nsusyf}) will have to  be of order $O(1/\delta)$ \footnote{
 For example to approximate
$1/\sqrt{2} =0.707106...$, to $n$ significant figures, $c,d$ would have to be $O(n)$.}.
Approximating,
\beq
\label{appns}
\alpha+\left|{c\over d}\right|\sim 2 \alpha,
\eeq
 we see that
\beq
\label{nsusyh}
{d^2 \over \alpha}\left(-\alpha^2+{c^2\over d^2}\right)\simeq 2 d^2\left(-\alpha + |{c\over d}|\right)
\sim O(1/\delta).
\eeq
It then follows  that  eq.(\ref{connonc}) will not be generically met,
since $\delta$ satisfies the condition, eq.(\ref{nsusyg}).

In other words, while $\alpha$ can be approximated arbitrarily well by the ratio of
two integers,
$|c/d|$, in general  doing so to better accuracy by choosing $\delta $ to be smaller
will make the condition, eq.(\ref{connonc}), harder to meet.

The condition in  eq.(\ref{connonc}) can be met
 if $\alpha$ is a non-generic irrational number
for which eq.(\ref{nsusyf}) can be met by taking
\beq
\label{nsusyj}
c,d \sim O\left({1\over \delta^{1/2-\epsilon}}\right).
\eeq
with $\epsilon>0$. In this case one finds that,
\beq
\label{nsusyk}
{d^2 \over \alpha}\left(-\alpha^2+{c^2\over d^2}\right)\sim O(\delta^{2\epsilon}),
\eeq
and thus eq.(\ref{connonc}) can be met if $\delta \ll 1$.

An example is provided by
\beq
\label{exnsusy1}
\alpha=\sqrt{p-1\over p}.
\eeq
It is easy to see that eq.(\ref{connonc}) is met in this case if $c=d=1$ and $p\gg 1$.
This example, fits in with the discussion above.
The irrational number $\alpha$, in this case,  is  well approximated
to $O(1/p)$ by two integers which are unity, and which therefore satisfies the condition,
eq.(\ref{nsusyj}).

The example above  can be easily generalised to the case,
\beq
\label{exnsusy2}
\alpha={m\over n} \sqrt{p-1 \over p}
\eeq
where $m< n$ and $mn \ll p$.  Once again eq.(\ref{connonc}) can be met, by taking,
$c=m,d=n$. We will have more to say about what these examples are teaching us in the following subsection,
where we consider varying the charges.

To summarise the discussion above,
we have  learned that eq.(\ref{connonc}) can be met,
but only for rather special values of  the initial charges.
These charges are such that $\alpha$ is of the form,
\beq
\label{nsusyl}
\alpha={m \over n}-\epsilon,
\eeq
where $0<\epsilon\ll 1$, and the integers, $m,n$ are not very big, and meet  the condition,
\beq
\label{nsusym}
2n^2\epsilon \ll 1.
\eeq
In this case, by taking, $c=m, d=n$ eq.(\ref{connonc}) can be met \footnote{For the matrix
eq.(\ref{elsl2z}) to exist $c,d$ must be coprime. This requires that we cancel off any common
factors in $m,n$ and take them to be coprime.}.

There is another way to characterise the non-genericity of $\alpha$. Suppose we
choose the initial charges such that $\alpha$ took a special value, eq.(\ref{nsusyl}),
and integers, $c,d$ exist meeting conditions, eq.(\ref{connonc}).
We could ask by how much can the initial charges be varied so that integers $c,d$
continue to exist, meeting the condition eq.(\ref{nsusyconnonc}).
If all the initial charges are of order $Q$ and they are varied by a small amount $\Delta Q$,
 we have that,
\beq
\label{chalpha}
{\Delta \alpha \over \alpha}\sim {\Delta Q\over Q}.
\eeq
Using, eq.(\ref{appns}), we can write the condition, eq.(\ref{nsusyconnonc}) as,
\beq
\label{snconc}
0<{d^2\over \alpha}\left(-\alpha^2+{c^2\over d^2}\right) \simeq 2d^2 \left(-\alpha + {c\over d}\right)<\delta.
\eeq
Now, when
\beq
\label{sncct}
\Delta \alpha  \sim {\delta \over 2 d^2},
\eeq
 $c,d$ will have to change from their initial
values for, the inequality, eq.(\ref{snconc}) to continue to hold. But for a
generic small variation,
new integers, $c,d,$ cannot be found meeting condition, eq.(\ref{nsusyj}),
rather the new integers will
be of order $O(1/\delta)$ and as a result eq.(\ref{nsusyconnonc}) will not be met.
Therefore  the maximum variation for the initial charges is of order,
\beq
\label{maxvar}
{\Delta Q \over Q} \sim {\delta \over 2 d^2}.
\eeq
Since $\delta$ satisfies eq.(\ref{nsusyg}), and $d$ is a non-vanishing integer,
 we see that this variation is small.

To summarise, in this subsection we have seen that a non-supersymmetric system
carrying generic $D0-D4$ brane charges cannot be brought to the Cardy limit after a
duality transformation.
The case when $\alpha$ is rational needs to be treated somewhat differently, we analyse this
case below.
Some examples, of non-generic charges, which can be brought to the Cardy limit using the duality
symmetry are discussed in appendix~B.
\subsection{Rational $\alpha$}
Since we saw that $\alpha$ had to be close to a rational number for the integers $c,d$ to exist
meeting the condition in  eq.(\ref{connonc}), it might seem at first
that for any  $\alpha$ which is  rational one can
always meet this condition. We show here that this is not true, eq.(\ref{connonc})
 can be met by rational $\alpha$ but again of a rather special form.

Suppose that
\beq
\label{ralpha}
\alpha={m\over n}
\eeq
so that $\epsilon$ in eq.(\ref{nsusyl}) vanishes.  We will again take the case
where
$\alpha<1, \alpha \not\ll 1$ \footnote{ We impose this restriction since
if $\alpha \ll 1$, the charges are   be non-generic
to start with.}.
 Without loss of generality, we can take
$m,n$ to be co-prime.
One could now choose $d=m,c=n$ so that
\beq
\label{rconda}
\left|{d\over c}\right| -\alpha =0 \ll 1.
\eeq
However in this case we see that eq.(\ref{connonc}) is not met  at the other end,
since, $(|{d\over c}|-\alpha) \not> 0$.

We need to find integers, $c,d$
 such that $|{d\over c}|$ is close to $\alpha$,
but does not exactly cancel it. This will not be generically possible for exactly
the same reason as the case of irrational $\alpha$. To meet the condition eq.(\ref{nsusyf}),
 $c,d$ will generically be  of order $1/\delta$ , while to meet eq.(\ref{connonc})
 they would  need to meet condition
eq.(\ref{nsusyj}). These two requirements are not compatible.

To understand when the condition in  eq.(\ref{connonc}) can be met more precisely,
let us write this equation  as,
\beq
\label{rec}
0<{1\over \alpha}(\alpha |d|+|c|)(-\alpha |d|+|c|) \ll 1.
\eeq
Now since, $ |c| > \alpha |d| $ we have, $|c|+|d| \alpha > 2 |d| \alpha$, and
it follows from eq.(\ref{rec}) that,
\beq
\label{red}
0<{2  |d| \over n} (n|c|-m|d|) \ll 1.
\eeq
Since, the minimum non-vanishing value of $(n|c|-m |d|)$ is unity, one consequence of
eq.(\ref{red}) is that,  $n/|d|\gg 1$. Given that   $\alpha$ is not much smaller than unity
it follows then that,
\beq
\label{ree}
m,n \gg 1.
\eeq
Also since, $2|d|>1$, it follows from eq.(\ref{red}) that
\beq
\label{reg}
0<{n|c|-m|d| \over n} \ll 1.
\eeq

In summary, if $\alpha$ is a rational number, $\alpha=m/n$,
an $SL(2,\mathbb Z)$ transformation can be found bringing the charges to a form
where condition, eq.(\ref{connonc}) is met,
 if two integers, $c,d$ exist  which are coprime, and which satisfy
the condition, eq.(\ref{red}). Generically, we have argued above,
  such integers do not exist,
and thus eq.(\ref{connonc}) will not be met.

One final comment before we move on. In the analysis above we
considered the case where $\vec{Q}_m'^2$ had the same sign as $\vec{Q}_e^2$.
If instead $\vec{Q}_m'^2$ has the opposite sign as $\vec{Q}_e^2$, the condition,
eq.(\ref{relnonb}) is replaced by,
\beq
\label{relnonbalternate}
0<d^2\alpha -{c^2 \over \alpha} \ll 1.
\eeq
The discussion above, for the irrational and rational values of $\alpha$, then
goes through essentially unchanged leading to similar conclusions. For generic
values of the charges, condition eq.(\ref{relnona}) will not be satisfied.
The condition in  eq.(\ref{nsusyl})
in this case is replaced by the requirement that
\beq
\label{nccase2}
\alpha={m\over n}+\epsilon,
\eeq
with $\epsilon>0$, such that,
\beq
\label{nccase3}
2n^2\epsilon \ll 1.
\eeq
If this requirement is met, eq.(\ref{relnonbalternate}) can be met by taking, $c=n, d=m$.
For rational, $\alpha$,  eq.(\ref{red}) is replaced by,
\beq
\label{caltc}
0<{2|c| \over m}(m|d|-n|c|) \ll 1.
\eeq

\subsection{Changing  The Charges}

In our discussion above for the non-supersymmetric case
we saw that for rather special values of $\alpha$ the condition,
eq.(\ref{connonc}) can be met. An example is given in eq.(\ref{exnsusy2}).
This prompts one to ask the following question: Although a generic charge configuration cannot be
brought to the Cardy limit, can we find  a
charge configuration lying near by, which can be brought to the Cardy limit ?
In this subsection we will answer the question. For large charges, $Q \gg 1$, we show that
such a near-by charge configuration does exist in the non-supersymmetric case. In contrast,
in the supersymmetric case, such a near-by configuration does not exist.

Before proceeding let us state more clearly what we mean by a   charge configuration lying
near  the starting $D0-D4$ configuration.   Suppose we carry out a change in the charges,
\beqa
\label{change}
\vec{Q}_e & \rightarrow &  \vec{Q}_e+\Delta \vec{Q}_e \\
\vec{Q}_m & \rightarrow & \vec{Q}_m + \Delta \vec{Q}_m.
\eeqa
The change is small, and the new charge configuration is near the original one,  if the conditions,
\beqa
\label{csmall}
\left|{\vec{Q}_e \cdot \Delta \vec{Q}_{e,m}  \over (\vec{Q}_{e,m})^2}\right| &  \ll &  1 \cr
\left|{\vec{Q}_m \cdot \Delta\vec{Q}_{e,m} \over (\vec{Q}_{e,m})^2}\right| & \ll & 1 \cr
\left|{\Delta \vec{Q}_{e,m} \cdot \Delta \vec{Q}_{e,m} \over (\vec{Q}_{e,m})^2}\right| & \ll & 1,
\eeqa
are met \footnote{These conditions are manifestly invariant under the $O(6,22,\mathbb Z)$ group.
Once we choose a particular basis to write the initial
 charges as, $(\vec{Q}_e, \vec{Q}_m)$, there is no residual
$SL(2,\mathbb Z)$ invariance left. The conditions, eq.(\ref{csmall}), are written in this basis, and are in-effect also $SL(2,\mathbb Z)$ invariant.}.
In these inequalities,  $\Delta  \vec{Q}_{e,m}$ in the numerator
 stands for either, $\Delta \vec{Q}_e$, or $ \Delta \vec{Q}_m$, the inequality holds in both cases.
 Similarly, $\vec{Q}_{e,m}$ in the denominator stands for either $\vec{Q}_e$ or $\vec{Q}_m$.
Note that it follows from these conditions that the change in the duality invariant, $I$,
eq.(\ref{definv}), and therefore also the change in the entropy, eq.(\ref{entsusy}),
eq.(\ref{entnonsusy}), is small.

Let us first consider  the supersymmetric case.
The required condition for an $SL(2,\mathbb Z)$ transformation, eq.(\ref{elsl2z}), to exist is that $\alpha$,
eq.(\ref{defalpha}),
satisfies the condition, eq.(\ref{csusya}). Suppose we start with generic  charges,
where $\alpha \le 1$, but
where  condition eq.(\ref{csusya}) is not met, and now carry out the  change in the charges,
eq.(\ref{change}).
The initial charges, $\vec{Q}_e, \vec{Q}_m,$ are both either space-like or time-like,
and since condition eq.(\ref{csusya}) is not met, are roughly comparable in magnitude.
It is then clear, and straightforward to verify explicitly,
 that small changes, meeting conditions, eq.(\ref{csmall}),
will not allow, eq.(\ref{condb}) to be met. We learn then that in the supersymmetric case there is no near by
 configuration - obtained by a small change in charges-  which brings the charges to the Cardy limit.

Next we come to the non-supersymmetric case.
Here one of the two vectors, $\vec{Q}_e, \vec{Q}_m$ is space-like and the other time-like,
and this makes the analysis more involved, as we have already seen above.
We will explicitly construct a  new set of charges, close to the original one and show
that it can be taken to the Cardy limit after a duality transformation.
The construction will be based on the example, eq.(\ref{exnsusy2}), and will proceed in two steps.
We will first find an altered set of charges
for which an $SL(2,\mathbb Z)$ transformation meeting condition, eq.(\ref{connonc}), exists.
Then in the second step we will
further  alter these charges so that the $SL(2,\mathbb Z)$ transformation we have identified in the first step,
followed by  an appropriate $O(6,22,\mathbb Z)$ transformation,
  brings this final set of altered charges to the Cardy limit.
 At both stages we will ensure that the changes in the charges are small
and that the conditions, eq.(\ref{csmall}), are met.

 In the starting configuration, the  $D0-D4$ brane charges   are large, of order, $Q$,
and roughly comparable, so that $\alpha$ satisfies condition, eq.(\ref{condaa}), but
$\alpha \not\ll 1$.

\noindent
{\bf  The First Step:}

In the first step, we then  change the $D0-D4$ charges (no new charges are excited at this stage)
 so that the new value of $\alpha$ is a rational, $m/n$.
The change in $\alpha$ can be kept small,
\beq
\label{conce}
\left|\alpha-{m\over n}\right|<\epsilon,
\eeq
with,
\beq
\label{condepsilon}
\epsilon <1,
\eeq
if we take the integers, $m,n$ to be sufficiently large,
\beq
\label{magmn}
m,n \sim  O(1/\epsilon).
\eeq
The required change in the charges is of order $\Delta Q$ where,
\beq
\label{ratq}
{\Delta Q \over Q} \sim {\Delta \alpha \over \alpha} \sim \epsilon
\eeq
Next, we change one of the $D4$-brane charges by order unity, this gives rise to a final value of\footnote{
For example, if only , $p^2,p^3\ne 0$, in the basis, eq.(\ref{defqm}),
then changing $p^2$ by unity would give,
$\alpha={m\over n}\sqrt{1-{1\over p^2}} ={m\over n}\sqrt{1-{1\over Q}},$ if $p^2=Q$.}
$\alpha$,
\beq
\label{falpha}
\alpha={m\over n} \sqrt{1-{1\over Q}}.
\eeq
Now choosing,
\beq
\label{defcd}
c=m, d=n,
\eeq
 eq.(\ref{nsusyconnonc}) is met, if the condition,
\beq
\label{ccc}
{m n \over Q } < \delta,
\eeq
is valid.
Using eq.(\ref{magmn}) this gives,
\beq
\label{ccd}
\epsilon > {1 \over \sqrt{\delta Q}}.
\eeq
We will see below, that $\delta$ which was introduced first in eq.(\ref{nsusyconnonc}),
can be taken to be a fixed small number, meeting condition, eq.(\ref{nsusyg}), and not scaling like an
inverse power of  $Q$.
Then by taking $Q$ to be sufficiently big, so that
\beq
\label{cce}
Q \gg {1\over \delta} \gg 1,
\eeq
 condition eq.(\ref{ccd}) can be made  compatible with eq.(\ref{condepsilon}).
To keep the shift in the charges small, it is best to take $\epsilon$ to be as small as possible,
subject to the condition, eq.(\ref{ccd}). We will take,
\beq
\label{satin}
\epsilon\sim {1\over  \sqrt{ \delta Q}}.
\eeq

It is useful in the subsequent discussion to distinguish between the altered  charges obtained
at this stage and the original charges we started with.  We denote the altered charges by the
tilde superscript. In the basis, eq.(\ref{defqe}), eq.(\ref{defqm}), we have,
\beqa
\label{tcharge}
\vec{\tilde{Q}}_e & = & (\tilde{q_0}, -\tilde{p}^1,0,0, \cdots, 0) \cr
\vec{\tilde{Q}}_m & = & (0,0,\tilde{p}^i,0,0,0, 0).
\eeqa

Before proceeding further it is worth examining condition eq.(\ref{ccc}) more carefully.
The inequality, eq.(\ref{connonc}), arose from eq.(\ref{condb}). It's stronger form
  is given by the condition  in eq.(\ref{condaalt}).
Here,  $p^{1'}$ is the charge that arised due to the $D4$-branes wrapping the K3,
in the final configuration which lies  in the Cardy limit and which is  obtained by starting
 with the altered charges and doing the duality transformation.
From eq.(\ref{condaalt}), eq.(\ref{nsusyconnonc}) we see that $\delta$ must satisfy the condition,
\beq
\label{condelta}
\delta \ll {1\over |p^{1'}|}.
\eeq
Now if $p^{1'}\sim Q$
we see that eq.(\ref{condelta}), eq.(\ref{ccd}), together imply that
 the condition  in eq.(\ref{condepsilon}) cannot be met.
We will see below that the final charge configuration has a value for $p^{1'}$ which is much smaller than
$Q$. In fact $p^{1'}$ can be taken to be $O(1)$ and not $O(Q)$. Thus, as was mentioned above,
$\delta$ can be taken to be  a small number not scaling like an inverse power of $Q$.
One can then choose $Q$ to meet the condition, eq.(\ref{cce}), and this will
then  suffice to meet eq.(\ref{ccd}) and   eq.(\ref{condepsilon}).

From eq.(\ref{ratq}) and eq.(\ref{satin}) we see that the
 required change in the charges are of the order,
\beq
\label{cch}
{\Delta Q \over Q} \sim \epsilon \sim {1\over \sqrt{ \delta Q}}.
\eeq
This gives,
\beq
\label{cci}
\Delta Q \sim   \sqrt{{Q \over \delta}}.
\eeq
We see that while,  $\Delta Q \gg 1$, from eq.(\ref{cch}), eq.(\ref{cce}), it follows
 that,
\beq
\label{ccj}
{\Delta Q \over Q} \sim {1\over \sqrt{ \delta Q}} \ll 1,
\eeq
so that the fractional change in the charges are small. Condition eq.(\ref{ccj})  ensures
that the requirements in eq.(\ref{csmall}) are met, so that the changes in charge are small.

We have now completed the first step.
The $SL(2,\mathbb Z)$ transformation that   takes  the  altered charges to the Cardy limit has the form,
\beq
\label{newsl}
A=\pmatrix{a & b \cr m & n \cr }
\eeq
The integers $m,n$ have been determined in terms of $\alpha$ for the altered charges above
eq.(\ref{falpha}). As discussed in  appendix B, $a,b,$ can  be chosen so that
they satisfy the conditions,
\beqa
\label{ornewab}
a& \sim &  O(m)  \nonumber \\
b & \sim & O(n).
\eeqa
The relations in eq.(\ref{ornewab}) will be important in the following discussion.

\noindent
{\bf The Second Step:  }

We now proceed to the second step and construct the $O(6,22,\mathbb Z)$ transformation.
This will require a further change in the charges. We will excite extra charges
which lie in the last two $\cal{H}\oplus \cal{H}$ subspaces in eq.(\ref{metric}).
 These are charges
which arises from the $T^2$.
The altered charges at the first stage are given in eq.(\ref{tcharge}). We now change them further,
so that the final altered charges take the form,
\beqa
\label{faltch}
\vec{\tilde{Q}}_e & = & (\tilde{q}_0,-\tilde{p}^1,0,0, \cdots, -b,0,n,0) \nonumber \\
\vec{\tilde{Q}}_m  & = & (0, 0, \tilde{p}^i, a, 0, -m,0).
\eeqa
Here $a, b,m,n $ are  elements of the $SL(2,\mathbb Z)$ matrix, eq.(\ref{newsl}).
Note that, $\tilde{q}_0, \tilde{p}^i \sim O(Q)$. From eq.(\ref{magmn}), eq.(\ref{ornewab}),
 we see that  $a,b,m,n \sim 1/\epsilon$. From, eq.(\ref{satin}) we then learn that
\beq
\label{finin}
a,b,m,n \sim  {1\over \epsilon} \sim \sqrt{ \delta Q}.
\eeq
The changes in charges that give eq.(\ref{faltch}) then meet the condition
\beq
\label{ccn}
{\Delta Q \over Q} \sim \sqrt{\delta \over Q} \ll 1,
\eeq
where the last inequality follows from the fact that the charge $Q$ meets  the condition,
eq.(\ref{cce}). This ensures that the conditions in eq.(\ref{csmall}) are met.

The $SL(2,\mathbb Z)$ transformation, eq.(\ref{newsl}), followed by an $O(6,22,\mathbb Z)$ transformation
that we  describe  explicitly in appendix C,  now brings the charges, eq.(\ref{faltch}) to the form,
\beqa
\label{finalcharge}
\vec{Q}_e' & = & (a\tilde{q}_0, 1, b \tilde{p}^i,   0, -ma\tilde{q}_0{\tilde p}^1, 1, -a \tilde{q}_0
 (a \tilde{p}^1+1) ) \cr
\vec{Q}_m' & = & (m \tilde{q}_0, 0, n\tilde{p}^i, 1, -m^2\tilde{q}_0\tilde{p}^1,0,
-m (a\tilde{p}^1+1)\tilde{q}_0).
\eeqa
These charges are in the Cardy limit. Since the second entry in $\vec{Q}_m'$ vanishes, the $D6$-brane
charge vanishes.   From the second entry in $\vec{Q}_e'$ we see that $|p^{1'}|$ is unity, as was promised above.
Finally, the extra charges excited in going from eq.(\ref{tcharge}) to eq.(\ref{faltch}) does not
change the value of $(\vec{\tilde{Q}}_m)^2$. Thus,
\beq
\label{magqmf}
{(\vec{Q}_m')^2 \over \sqrt{|I|}} \simeq \left({m n \over Q}\right) \simeq \delta \ll 1,
\eeq
where we have used eq.(\ref{magmn}) for $m,n$ and eq.(\ref{satin}) for $\epsilon$.
It then follows that  eq.(\ref{condaalt}) is met and the final charges are in the Cardy limit.

Two comments before we end. First, there is some leeway in the  $O(6,22, Z)$ transformation
 which acting on the charges, eq.(\ref{faltch}), brings them to the Cardy limit. For example,
an $O(6,22,\mathbb Z)$ transformation can be found that  results in $p^{1'}$ being a  number much large than
unity, but  not scaling with $Q$.
Second, we have seen in subsection 3.2 that in the vicinity of one set of charges which can
brought to the Cardy limit, are other near by  charges meeting condition, eq.(\ref{maxvar}), which can
also be taken to the Cardy limit. Using, eq.(\ref{defcd}), eq.(\ref{magmn}), we see that
eq.(\ref{maxvar}) takes the form,
\beq
\label{varc}
{\Delta Q \over Q} \sim \delta \epsilon^2.
\eeq
Since, $\delta \ll 1, \epsilon<1$,  the size of this variation,
${\Delta Q \over Q} \ll \epsilon$.
Thus starting from one of the special charge configurations which can be brought to the Cardy limit,
a variation of order, eq.(\ref{varc}), takes us to charges of the generic kind which can no longer
be taken to the Cardy limit by a duality transformation. These charges will have to be changed by
an amount of order, eq.(\ref{ratq}), to be able to bring them to the Cardy limit.

\section{The $D0-D6$ System}
In this section we consider the $D0-D6$ system,  where  only $q_0,p^0\ne0$, and all other charges vanish,
 eq.(\ref{defqe}), eq.(\ref{defqm}). We show that such a charge configuration can never be brought to the
Cardy limit.
For this set of charges we have the following relations,
\beqa
\label{rel06a}
\vec{Q}_e^2 & = & 0 \cr
\vec{Q}_m^2 & = & 0 \cr
\vec{Q}_e\cdot \vec{Q}_m& = & q_0p^0.
\eeqa
The invariant $I$, eq.(\ref{definv}), is,
\beq
\label{inv06}
I=-(q_0p^0)^2,
\eeq
It is   negative,  and the  state breaks supersymmetry.

Let us   assume that there is an $SL(2,\mathbb Z)$
transformation, eq.(\ref{elsl2z})
 which followed by an $O(6,22,\mathbb Z)$ transformation  brings the charges
to the Cardy limit.
Denoting the final charges by $\vec{Q}_e', \vec{Q}_m',$ we have that,
\beq
\label{valqmf}
\vec{Q}_m'^2=2 c d q_0 p^0.
\eeq
If the final charges are in the Cardy limit, it follows from eq.(\ref{two}), and the fact that
$|p^{1'}| \ge 1$ that,
\beq
\label{cd0d6}
{|(\vec{Q}_m')^2| \over \sqrt{|I|}}\ll 1.
\eeq
From, eq.(\ref{valqmf}) and eq.(\ref{inv06}), this leads to the condition,
\beq
\label{c06g}
|c d| \ll 1.
\eeq
Now note that $c,d$ are integers. Thus the only way in which eq.(\ref{c06g})
 can be met is if $c d=0$.
This will mean that  $\vec{Q}_m'^2=0$ and hence the central charge, eq.(\ref{cchargegen}),
for the final charges   vanishes.  We do not want the central charge to vanish since the
resulting $AdS_3$ space-time would not be described by weakly coupled supergravity.
As a result we find that there is no duality transformation which can bring the $D0-D6$
system to the Cardy limit.

In parallel with our discussion of section 3.4 we now ask if there are near by charges
which can be brought to the Cardy limit. The following construction shows that such a  set of
charges does exits, as in the
non-supersymmetric $D0-D4$ system.
The $D0-D6$ system we start with has charges which  in the basis, eq.(\ref{defqe}),
eq.(\ref{defqm}), are given by,
\begin {eqnarray}
\label{charged0d6}
\vec{Q}_e & = &   (q_0, 0, \cdots, 0)  \nonumber \\
\vec{Q}_m & = & (0,p^0, 0, \cdots, 0).
\end {eqnarray}
The charges meet the condition,
\beq
\label{condad0d6}
|\vec{Q}_e\cdot \vec{Q}_m| =|q_0p^0|\gg 1.
\eeq
For the change in the charges to be small the condition,
 analogous to eq.(\ref{csmall}) in the $D0-D4$ case, is given by,
\beqa
\label{cd0d6small}
\left|{\vec{Q}_e \cdot \Delta \vec{Q}_{e,m}
\over \vec{Q}_e\cdot \vec{Q}_m}\right| &  \ll &  1 \cr
\left|{\vec{Q}_m \cdot \Delta\vec{Q}_{e,m}
\over \vec{Q}_e \cdot \vec{Q}_m}\right| & \ll & 1 \cr
\left|{\Delta \vec{Q}_{e,m} \cdot \Delta \vec{Q}_{e,m}
\over \vec{Q}_e\cdot \vec{Q}_m}\right| & \ll & 1.
\eeqa

Now consider the altered  charges,
\beqa
\label{ad0d6}
\vec{Q}_e & = & (q_0,0,1,0, \cdots,0) \nonumber \\
\vec{Q}_m & = & (0,p^0,-1,1, \cdots, 0).
\eeqa
It is easy to see that conditions, eq.(\ref{cd0d6small}),
are met and the changes in the charges are small.

In eq.(\ref{ad0d6}), we have activated additional charges lying in the second Hyperbolic
sublattice, $\cal{H}$, defined in eq.(\ref{metric}). We could have instead
activated the additional
 charges to lie in any of the other  Hyperbolic sublattices  (or infact the
${\cal E}_8$ sublattices),
and a similar discussion would go through.

Now consider an $O(2,2)$ transformation acting on the two $\cal{H}$
sublattices in which the charges lie, of the form,
\beq
\label{o2206}
\pmatrix{1 & 0 & 0 & 0 \cr 0 & 1 & p^0 & 0 \cr
         0 & 0 & 1 & 0 \cr -p^0  & 0 & 0 & 1 \cr}.
\eeq
This  brings the altered  charges,  eq.(\ref{ad0d6}), to the form,
\beqa
\label{altd0d6}
\vec{Q}_e' & = & (q_0,p^0,1,-p^0 q_0, 0,  \cdots , 0 ) \nonumber \\
\vec{Q}_m'& =  & (0, 0, -1, 1, 0, \cdots 0).
\eeqa
These charges are in the Cardy limit. The second entry in $\vec{Q}_m'$ vanishes,
therefore, $p^{0'}=0$. Also,
$p^{1'}=p^0, (\vec{Q}_m')^2 =-2$,   so that the condition,
eq.(\ref{two}), is met, as long as
\beq
\label{condq0}
|q_0|\gg 1.
\eeq

Note that the central charge, $C \sim |p^{1'} (\vec{Q_m'})^2| \sim (p^0)^2$.
This meets the condition, $C\gg 1$ if $|p^0|\gg 1$.
Alternatively, if $p^0\sim O(1)$, we can excite additional charges in eq.(\ref{ad0d6})
so that, for example, $p^{1'} \gg 1$, and thus $C\gg 1$.

\section{Absence of Magnetic Monopole Charge}
 We have mentioned  above that lifting a configuration with  $D6$ brane charge to
M-theory cannot give a locally  $AdS_3$ spacetime in the near-horizon limit.
 We prove this statement here.

We start with a general extremal  black hole, carrying charges given in eq.(\ref{defqe}),
eq.(\ref{defqm}),   in four dimensions in IIA theory. The near horizon geometry is 
$AdS_2\times S^2$.  An $AdS_2$ space-time has $SO(2,1)$ symmetry.
This gets enhanced to $SO(2,2)$ in the $AdS_3$ case \footnote{Our analysis of the symmetries
in this section will be local. So the breaking of $SO(2,2)$ symmetry
 due to identifications which are made  in the BTZ geometry
will not be relevant.}. In the special case where 
 the black hole carries no $D0$-brane charge,  $N$ units of $D6$-brane charge, and arbitray 
values of the other charges,  it is
well known that one does not get the $SO(2,2)$ symmetry
of $AdS_3$ in the near horizon limit geometry.
 The $D6$-brane charge is KK monopole charge  along the M direction.
This  charge results in the M-direction being fibered over the
$S^2$ resulting in the near horizon geometry of form, $AdS_2 \times  S^3/Z_N$.

Here we will examine what happens if the black hole carries both $D0$ and $D6$ brane charges,
besides having arbitrary values of the other charges, 
and find that the symmetries of the near horizon geometry are
$SO(2,1) \times SO(3) \times U(1)$
and are  therefore  not enhanced to $SO(2,2)$.
This proves that the only way to get a locally $AdS_3$ geometry on lifting to M-theory
is for the $D6$-brane charge to vanish.

Lifting the $AdS_2\times S^2$ near-horizon geometry to M-theory, gives,
\beqa
\label{5dnh}
ds^2 &=& R^2(-\cosh^2\theta_1 d\phi_1^2  + d\theta_1^2) +
R^2(d\theta_2^2+\sin^2\theta_2 d\phi_2^2) \cr
&+& g_{\psi\psi}(d\psi+\alpha \sinh\theta_1 d\phi_1 + \beta \cos\theta_2 d\phi_2)^2
\eeqa
Here we are using Global coordinates  $\theta_1, \phi_1$ for  $AdS_2$,
 polar coordinates, $\theta_2,\phi_2$
for the $S^2$, and denoting  the M-theory direction as $\psi$.
The metric component, $g_{\psi\psi}$, is a constant.
 $\alpha, \beta$ are proportional
to the $D0$ and $D6$ brane charges and are non-vanishing if these charges are non-vanishing.
We seek the Killing vectors for this metric.

It is convenient to analytically continue the $AdS_2$ metric to that of $S^2$ as follows,
\beqa
\label{ac}
\theta_1 & \rightarrow &  i \left({\pi\over 2}- \theta_1\right) \cr
(R^2)_{AdS}  & \rightarrow &  -R^2 \cr
\alpha  & \rightarrow  & - i \alpha .
\eeqa
This gives,
\beqa
\label{5dac}
ds^2&=& R^2(d\theta_1^2+\sin^2\theta_1 d\phi_1^2) + R^2(d\theta_2^2+\sin^2\theta_2 d\phi_2^2)
\cr &+& g_{\psi\psi}(d\psi+\alpha \cos\theta_1 d\phi_1 + \beta \cos\theta_2 d\phi_2)^2.
\eeqa
We show that the isometry group   of this metric is,  $SO(3) \times SO(3) \times U(1)$,
it will then follow by analytic continuation that the isometry group  of eq.(\ref{5dnh})
is,  $SO(2,1) \times SO(3) \times U(1)$.

By rescaling the $\psi$ coordinate, $\alpha$ and $\beta$, this metric can be written as,
\beqa
\label{nmet}
ds^2&=& R^2[(d\theta_1^2+\sin^2\theta_1 d\phi_1^2) + (d\theta_2^2+\sin^2\theta_2 d\phi_2^2)
\cr &+&  (d\psi' + \alpha' \cos \theta_1 d \phi_1 + \beta' \cos \theta_2 d\phi_2)^2].
\eeqa
$\alpha', \beta'$ are proportional to $\alpha, \beta$ and only vanish when the latter do.
Next we drop the overall factor of $R^2$, and rescale $\phi_1, \phi_2$ as follows,
\beq
\label{rp}
\alpha '\phi_1\rightarrow  \phi_1, \ \ \  \beta' \phi_2 \rightarrow \phi_2.
\eeq
Note this rescaling is well defined only if $\alpha', \beta',$ and hence $\alpha, \beta,$ are
non-vanishing.
This gives for the metric,
\beqa
\label{newmb}
ds^2 & = & d\theta_1^2 + d\theta_2^2 + (1+ (\tilde{\alpha})^2 \sin^2\theta_1 ) d\phi_1^2
 + (1+(\tilde{\beta})^2 \sin^2\theta_2) d\phi_2^2  +d\psi^2  \cr
          &&  + 2 \cos\theta_1d\psi d\phi_1
+ 2 \cos\theta_2 d\psi d\phi_2 + 2 \cos\theta_1 \cos\theta_2 d\phi_1 d\phi_2,
\eeqa
where,
\beqa
\label{deftab}
({\tilde \alpha})^2 & = &  {1\over \alpha^{'2}}-1 \\
(\tilde{\beta})^2 & = & {1\over \beta^{'2}}-1.
\eeqa
To save clutter we will henceforth drop the tildes on $\alpha, \beta$ and denote
the metric in eq.(\ref{newmb}) as,
\beqa
\label{newmc}
ds^2 & = & d\theta_1^2 + d\theta_2^2 + (1+ \alpha^2 \sin^2\theta_1 ) d\phi_1^2
 + (1+\beta^2 \sin^2\theta_2) d\phi_2^2+d\psi^2 \cr
&& + 2 \cos\theta_1d\psi d\phi_1
+ 2 \cos\theta_2 d\psi d\phi_2 + 2 \cos\theta_1 \cos\theta_2 d\phi_1 d\phi_2.
\eeqa
The reader should note that $\alpha, \beta,$ in eq.(\ref{newmc}) are different from
$\alpha, \beta,$ as appearing in eq.(\ref{5dac}).

We now turn to studying the isometries of the metric, eq.(\ref{newmc}).
First note that $\partial_{\phi_1}, \partial_{\phi_2}, \partial_\psi,$ are commuting isometries
of this metric. They can be taken to be part of the Cartan generators of the full isometry
group. Any other killing vector,  $\xi$, can then be taken to  carry definite charges with
 respect to these
generators, and   satisfies the relations,
\beq
\label{relc}
[\partial_{\phi_1},\xi]  =   i m_1 \xi,
\eeq
\beq
\label{relcb}
[\partial_{\phi_2},\xi]  =   i m_2 \xi,
\eeq
\beq
\label{relcd}
[\partial_\psi,\xi]  =   i m_3 \xi,
\eeq
where $m_1,m_2,m_3$ are the  eigenvalues with respect to these three isometries.

The killing  vector,  $\xi$, must satisfy the Killing  conditions,
\beq
\label{kcond}
\partial_\alpha \xi^\gamma g_{\gamma \beta} + \partial_\beta \xi^\gamma g_{\gamma \alpha}
+\xi^\gamma \partial_\gamma g_{\alpha \beta}=0
\eeq
for all values of $\alpha, \beta$.

These Killing conditions are studied in more detail in appendix D.
 One finds
that there are only four more non-trivial Killing vectors, corresponding to
$m_1 =\pm \sqrt{1+\alpha^2}, m_2=m_3=0$ and $m_2= \pm \sqrt{1+\beta^2}, m_1,m_3=0$.
Altogether there are then seven  Killing vectors, given by,
\beqa
\label{kvectors}
\xi_1 & = & e^{i\sqrt{1+\alpha^2}\phi_1}\left[\partial_{\theta_1}+
{i\over \sqrt{1+\alpha^2}} \cot\theta_1  \partial_{\phi_1}
-{i\over\sqrt{1+\alpha^2}}{ 1 \over \sin\theta_1} \partial_\psi\right]  \cr
\xi_2 & = & e^{-i\sqrt{1+\alpha^2} \phi_1}\left[\partial_{\theta_1}-
{i\over \sqrt{1+\alpha^2}} \cot\theta_1  \partial_{\phi_1}
+{i\over\sqrt{1+\alpha^2}}{1\over \sin\theta_1}\partial_\psi\right] \cr
\xi_3 & = & \partial_{\phi_1}  \cr
\xi_4 & = & e^{i\sqrt{1+\beta^2} \phi_2}\left[\partial_{\theta_2}+
{i\over \sqrt{1+\beta^2}} \cot\theta_2  \partial_{\phi_2}
-{i\over\sqrt{1+\beta^2}}{1\over \sin\theta_2}\partial_\psi\right]  \cr
\xi_5 & = & e^{-i\sqrt{1+\beta^2} \phi_2}\left[\partial_{\theta_2}-
{i\over \sqrt{1+\beta^2}} \cot\theta_2  \partial_{\phi_2}
+{i\over\sqrt{1+\beta^2}}{1\over \sin\theta_2}\partial_\psi\right] \cr
\xi_6 & = & \partial_{\phi_2} \cr
\xi_7 & = & \partial_\psi
\eeqa
The first three give rise to an $SO(3)$ isometry, the second three to another
$SO(3)$ and the last to an $U(1)$ isometry, giving the total symmetry group, $SO(3)\times
SO(3) \times U(1)$.
After analytic continuation this implies that the metric we started with has isometries,
$SO(2,1)\times SO(3)\times U(1)$.

We refer the reader to appendix D for more details.

\section{Conclusions}

This paper has two main results.
First, we have shown that a generic supersymmetric or non-supersymmetric system of charges
cannot be brought to the Cardy limit using the duality symmetries.
Second, we have found that the required non-genericity to be able to bring a set of charges
to  the Cardy limit is interestingly different in the supersymmetric and the non-supersymmetric
cases. For large charge, in the non-supersymmetric case but not the supersymmetric one,
we can always find a set of charges lying close by which can be brought to the
Cardy limit. The required shift in the charges satisfy the condition \footnote{More correctly,
the condition in the $D0-D4$ case is given in eq.(\ref{cch}),
where $\delta$ is a small number that does
not scale with $Q$, and in the $D0-D6$ case, with $q_0,p^0\gg 1$, it  is given by,
 ${\Delta Q \over Q}\sim {1\over Q}$.},
\beq
\label{resshift}
{\Delta Q \over Q} \sim {1\over \sqrt{Q}}.
\eeq
These results were proved for the $D0-D4$ system and the $D0-D6$ system. We expect
them to be more general.

For example, our analysis of the $D0-D4$ system,
   leading to the conclusion that generic charges cannot be brought
to the Cardy limit, immediately applies to all charges which satisfy the condition,
\beq
\label{concconda}
\vec{Q}_e\cdot \vec{Q}_m=0.
\eeq
Similarly, the analysis of the $D0-D6$ system
applies to all charges meeting the condition,
\beq
\label{conccondb}
\vec{Q}_e^2=\vec{Q}_m^2=0.
\eeq
with the conclusion that all such charges can never be brought to the Cardy limit.
Also, all the results immediately  apply to  other charges which lie in the same
 duality orbit  as  the D0-D4 or D0-D6 systems.

In our analysis we did not determine all the necessary and sufficient conditions
that need to be met to be able to bring a set of charges to the Cardy limit.
To obtain a more complete understanding of these conditions, for a general set of charges,
it would be useful to start with a classification
of all the  discrete invariants of $SL(2,\mathbb Z)\times O(6,22,\mathbb Z)$. It should be possible
to express the required
 conditions,  for any charge configuration to be brought to the Cardy limit,
 in terms of these invariants.  We leave such an analysis for the future.

Another approach would be to bring the charges to a  canonical form and
then carry out the analysis for general charges of this form.  As long as the charges lie in the
$\Gamma^{(6,6)}$ sublattice, made out of the $6$ Hyperbolic sublattices, $\cal{H}$ in
eq.(\ref{metric}), one can show  using the duality symmetries that  the electric
charges, $\vec{Q}_e$, can always be made to lie only in first hyperbolic sublattice, while the
magnetic charges, $\vec{Q}_m$, take non-trivial values  in the first two hyperbolic sublattices.
These results are discussed in appendix E. One expects these results to be further generalised,
when charges lying in the ${\cal E}_8 \times {\cal E}_8$ sublattice are also excited.
For example, it has shown that  a general  time-like vector can always be made to lie in
one Hyperbolic sublattice, (see  the discussion in
\footnote{Also, V.V.Nikulin, Math.USSR Izvestija,14(1980),pg.103.} \cite{hep-th/9810210}).
Further analysis along these lines is also left for the future.

Our conclusions in the  supersymmetric  case are  in accord with   recent results obtained
for the subleading corrections to the entropy, going like    $1/Q$.
If the system could be brought
to the Cardy limit these corrections  would be of the form, eq.(\ref{cardyent}), with
the central charge receiving
$1/Q$ corrections. The results for the first subleading corrections, which have been obtained by
directly counting the dyonic degeneracy and computing the four derivative corrections using
the Gauss-Bonnet term, are now known not to be  generally of this form, \cite{hep-th/0412287},
\cite{Sen:2005iz}, \cite{0605210}.

One of the main motivations of this investigation was to ask how far the $AdS_3/CFT$ description
can take us in understanding the entropy of non-supersymmetric black holes.
If the charges lie in the Cardy limit, then at least in some region of moduli space, the black
hole with these charges can be viewed as a BTZ black hole in $AdS_3$ space.
The microscopic states which account for the black hole entropy can then be understood as states
in a $1+1$ dim. CFT, and their entropy can be easily found in terms of the Cardy formula.
Our result, that in the non-supersymmetric case a generic set of charges, after a small shift,  can be brought to the Cardy limit is quite promising in this context. It tells us that such a microscopic counting for the leading order entropy is available for generic charges, at least in some region of moduli space.

The main complication in determining the entropy microscopically is then it's possible moduli dependence.
This is a particularly important issue in the non-supersymmetric case. In the Cardy formula the entropy is determined by the central charge.   Now,
the central charge is  protected by anomaly considerations and is therefore moduli independent.
Thus for  the charges which can be brought to the Cardy limit, the entropy must be
moduli independent, at least for small shifts of moduli \footnote{Larger shifts might result
in a jump, akin to a phase transition, where  the formula for the entropy gets significant
corrections.}.
 Since the required fractional shift
to get to such a configuration is small, of order,
 $O(1/\sqrt{Q})$, eq.(\ref{resshift}), one would hope that
this is enough to prove that the leading entropy is generally
  moduli independent.

Once the moduli independence of the entropy is established, it is
 easy to furnish  an argument, as follows,  leading to the determination of the
entropy microscopically.
The entropy must now be  a function only of the charges. And the dependence on the charges
must enter through   invariants  of  the discrete duality group,
which is an exact symmetry of string theory.
For the case we are studying
here, one of these invariants, $I$, eq.(\ref{definv}), is also an invariant of the full continuous group,
$SL(2,{\mathbb R}) \times O(6,22,{\mathbb R})$.
The others are discrete invariants.  Now the discrete invariants are not continuous functions
of charge and typically  undergo big jumps when the charges are changed only
slightly \footnote{For example consider the discrete invariant,
gcd$(Q_e^iQ_m^j-Q_e^jQ_m^i,Q_e^kQ_m^l-Q_e^lQ_m^k), \forall i,j,k,l \in \{1,2, \cdots,28 \}$.
Since the gcd can vary discontinuously, this invariant can change by big jumps.}
It is  physically reasonable to demand that for large charges the leading order
 entropy  does not undergo such discontinuous jumps.
 This would mean that any dependence on the
discrete invariants must be subdominant at large charge \footnote{This argument was given
to us by  Shiraz Minwalla, we thank him for the discussion on this point and related issues.}.
The resulting functional dependence on the continuous invariant
can then be determined by taking any convenient set of charges, which gives rise
to a non-vanishing value for this invariant.
In particular one can always find charges in the Cardy limit for which this invariant does
not vanish.  For such a set of  charges a microscopic
calculation of the entropy is often possible as was mentioned above, and this
 would then determine the entropy for all general charges.

These arguments should also apply when one includes angular momentum in four dimensions, $\vec{J}$.
In this case there are  now two invariants of the continuous duality symmetries, and
 the Rotation group,  $I$ and
$\vec{J}^2$.   An argument along the above lines would fix the dependence on both these invariants.
Note that the resulting expression for the entropy would then also be valid when $I$,
and more generally all the charges, $\vec{Q}_e, \vec{Q}_m$ vanish, leading to microscopic
determination of the entropy of  an extreme Kerr black hole in four dimensions.
It is easy to check that the resulting answer is in agreement with the Beckenstein-Hawking
entropy in this case.

These arguments will be developed, at more length and with more care, in a forthcoming paper.

The arguments above, whose purpose is to provide a microscopic understanding of the entropy,
are already known to have counterparts on the gravity side.  This makes us hopeful that they can
be more fully fleshed out on the microscopic side as well. We end with a brief discussion
of these issues from the gravity point of view.

Recent advances have now established that the attractor mechanism is valid for
all extremal black holes, supersymmetric as well as non-supersymmetric ones (See
\cite{Ferrara:1995ih}, \cite{Strominger:1996kf}, \cite{hep-th/9602136}, \cite{hep-th/9603090},
 for early work.
More recent advances are in, e.g,  \cite{Sen:2005wa},\cite{hep-th/0507096}, \cite{Sen:2005iz},
\cite{hep-th/0511117}, \cite{Kallosh:2005ax},\cite{Giryavets:2005nf}, \cite{Goldstein:2005rr},
\cite{Kallosh:2006bt}, \cite{Kallosh:2006bx}, \cite{Kaura:2006mv}, \cite{hep-th/0603247},
\cite{Ferrara:2006xx}, \cite{Bellucci:2006xz}, \cite{Bellucci:2006ib},
\cite{Andrianopoli:2006ub}, \cite{D'Auria:2007ev}, \cite{Bellucci:2007gb},
\cite{Andrianopoli:2007rm}, \cite{Ferrara:2007pc}, \cite{Ferrara:2007tu},
\cite{Ceresole:2007rq}, \cite{Andrianopoli:2007kz}, see also,  \cite{Sen:2007qy},
and references therein).
This shows that the entropy is not dependent on the moduli \footnote{More correctly this shows that the
entropy is independent of small shifts in the moduli. There can be discontinuous jumps in the
entropy as the moduli
are varied, see ref Moore and Denef for related recent developments. However, this might be less
of a worry if we are interested in the entropy of a single-centered black hole.}.
Once the moduli independence is established the duality symmetries allow the entropy for
general charges to be related to the entropy which arise for
a set of  charges in the Cardy limit.
In the supergravity approximation, which is valid at large charge, the duality group is
enhanced to the full continuous group,  in the case we are considering here to
$SL(2,\mathbb R) \times O(6,22,\mathbb R)$.  A  duality transformation will act on both the charges and the moduli,
and to begin with the entropy could have been a duality invariant function of the moduli and charges.
However, once we have established that the entropy is moduli independent it must be an invariant
of the charges alone. Since there is only one duality invariant of the continuous group \footnote{We
are neglecting angular momentum, $\vec{J}$, here.}, $I$,
the entropy for a general set of charges can be related to the entropy for charges
 in the Cardy limit, with the same
value of this invariant.

\subsection*{Acknowledgements}
We would like to thank Atish Dabholkar, Per Kraus,  Shiraz Minwalla, Sunil Mukhi,
Arvind Nair, Suvrat Raju and  Ashoke Sen  for discussion.
   This research is supported
by the Government of India. S.P.T. acknowledges support from the Swarnajayanti Fellowship, DST, Govt.
of India. P.K.T. acknowledges support form the IC\&SR (IITM) Project No. PHY/06-07/157/NFSC/PRAS. S.N. is grateful to Atish Dabholkar and to LPTHE where some part of this work was done. S.N. acknowledges support from the Sarojini Damodaran Fellowship.
Most of all we thank the people of India for generously supporting research in String Theory.

\appendix
\section{Tightening the Conditions in the Supersymmetric Case}
A supersymmetric $D0-D4$ system, which can be taken to the Cardy limit, must meet the condition,
eq.(\ref{condfe}). In this appendix we show that this condition can be somewhat strengthened,
leading to eq.(\ref{sforma}).

This comes about as follows. In general the $SL(2,\mathbb Z)$ transformation, eq.(\ref{slfirst}),
will be followed by an $O(6,22,\mathbb Z)$ transformation, $B \in O(6,22,\mathbb Z)$, to obtain the final
configuration, $(\vec{Q}_e', \vec{Q}_m')$ which is given by,
\beqa
\label{fformq}
\vec{Q'_e} &= & B \vec{Q}_e + b B \vec{Q}_m \\
\vec{Q}'_m & = & B \vec{Q}_m.
\eeqa

 We will see shortly that this final configuration is in the
Cardy limit if and only if the configuration, $(\vec{\tilde{Q}}_e, \vec{\tilde{Q}}_m)$,
defined by,
\beq
\label{deftil}
(\vec{\tilde{Q}}_e, \vec{\tilde{Q}}_m)=(B \vec{Q}_e, B \vec{Q}_m)
\eeq
is in the Cardy limit.
Note that the charges, $(\vec{\tilde{Q}}_e, \vec{\tilde{Q}}_m)$, are
obtained by  applying only the transformation, $B \in O(6,22,\mathbb Z)$  on $(\vec{Q}_e,\vec{Q}_m)$.
Applying condition eq.(\ref{conda}) to the charges, $(\vec{\tilde{Q}}_e, \vec{\tilde{Q}}_m)$,
we learn that   for them to be in the Cardy limit,
\beq
\label{tildetwo}
|I|\gg \left(\tilde{p}^1 \left(\vec{\tilde{Q}}_m\right)^2\right)^2.
\eeq
From eq.(\ref{deftil}) we see that $\left(\vec{\tilde{Q}}_m\right)^2=\vec{Q}_m^2.$
Now since $\vec{\tilde{Q}}_e$ is obtained by applying an $O(6,22,\mathbb Z)$ transformation to $\vec{Q}_e$,
the minimum value $\tilde{p}^1$ can take is ${\rm gcd}(\vec{Q}_e)$. Eq.(\ref{sforma}) then follows,
after using eq.(\ref{valI}) for $I$.

To complete the argument let us show that $(\vec{Q}_e', \vec{Q}_m')$ can be in the Cardy limit if an only if
$(\vec{\tilde{Q}}_e, \vec{\tilde{Q}}_m)$ is in the Cardy limit.
To see this we note that from eq.(\ref{fformq}) and eq.(\ref{deftil})
 it follows that,
\beq
\label{relpta}
\vec{Q}_e'   = \vec{\tilde{Q}}_e+b \vec{\tilde{Q}}_m,
\eeq
and,
\beq
\label{relptb}
\vec{Q}_m' =\vec{\tilde{Q}}_m.
\eeq
If $\vec{Q}_m'$ is in the Cardy limit the $D6$-brane charge for this configuration must vanish,
so, $p^{0'}=0$. From eq.(\ref{relptb}) we see this implies that $\tilde{p}^0$
also vanishes. Eq.(\ref{relptb}) also implies that $(\vec{Q}_m')^2=(\vec{\tilde{Q}}_m)^2$.
And eq.(\ref{relpta}) implies that  $p^{1'}=\tilde{p}^1$.
The second condition for the Cardy limit, eq.(\ref{two}), is
\beq
\label{condin}
I\gg 6 (p_1'\vec{Q'}_m^2)^2.
\eeq
Since
 $I$ is a duality invariant, it then follows
that the condition eq.(\ref{condin}) is  the same as the corresponding condition in terms of the
tilde variables,
\beq
\label{condtila}
I\gg 6 \left(\tilde{p_1}\left(\vec{\tilde{Q}}_m\right)^2\right)^2.
\eeq

\section{Some Non-supersymmetric Examples}
In this appendix we present some examples of charges in th non-supersymmetric case, which can
be brought to the Cardy limit after a duality transformation.

We take,
\beq
\label{nsusye1}
\vec{Q}_e=(p-1,-1,0,0,0, \cdots 0)
\eeq
\beq
\label{nsusye2}
\vec{Q}_m=(0,0,1,p,0,\cdots 0),
\eeq
with,
\beq
\label{condp}
p\gg 1.
\eeq
The quartic invariant, $I$, eq.(\ref{definv})  is,
\beq
\label{valnonsusyexI}
I=-4p(p-1).
\eeq
The value of $p^1=1$, and $\vec{Q}_m^2=2p$, so we see that condition, eq.(\ref{two})
 is not met and the
starting configuration is not in the Cardy limit.
In this example, $|\vec{Q}_e^2|<|\vec{Q}_m^2|$, so that $\alpha>1$ to begin, we therefore
carry out the $SL(2,\mathbb Z)$ transformation, $\pmatrix{0&1\cr-1&0}$,
which gives,
\beqa
\label{nsusye3}
\vec{Q}_e & = & (0,0,1,p,0,\cdots) \\
\vec{Q}_m & = & -(p-1,-1,0,0,0,\cdots, 0).
\eeqa
The resulting value of $\alpha$ is,
\beq
\label{resal}
\alpha=\sqrt{p-1\over p}.
\eeq
This is of the form discussed above in eq.(\ref{exnsusy1}).
Starting with the charges, eq.(\ref{nsusye3}), we now carry out $SL(2,\mathbb Z)\times O(6,22,\mathbb Z)$
 transformations which bring it in the Cardy limit.
The  $SL(2,\mathbb Z)$ transformation is,
\beq
\label{elns}
A=\pmatrix{(p-1)&- p\cr1&- 1}
\eeq
with resulting charges,
\beqa
\label{nsusye5}
\vec{\tilde{Q}_e} & = & (p(p-1),-p,p-1,(p-1)p, 0 \cdots, 0) \\
\vec{\tilde{Q}_m} & = & (p-1,-1,1,p,0,\cdots,0)
\eeqa
This is followed by an $O(6,22,\mathbb Z)$ transformation,
\beq
\label{vaonse}
B=\pmatrix{1& 0 & 0& 0\cr
           0 & 1 & 1 & 0 \cr
           0 & 0 & 1 & 0 \cr
           -1 & 0 & 0 & 1 \cr
}
\eeq
By this we mean that $B$ acts non-trivially on the 4 dimensional sublattice of charges
where the inner product is given by the first two factors of $\cal{H}$ in eq.(\ref{inpro}), and acts
trivially on the rest of the lattice.
The transformation $B$ gives  the final charges,
\beqa
\label{fcnsex1}
\vec{Q}_e' &  = & (p(p-1),-1,p-1,0, 0, \cdots, 0) \\
\vec{Q}_m' & = & (p-1,0,1,1,0,\cdots 0).
\eeqa
Since the second entry in $\vec{Q}_m'$ vanishes, the  $D6$ brane charge in the final configuration vanishes as is needed for the Cardy limit.
From the second entry in $\vec{Q}_e'$ we see that $|p^{1'}|=1$, and we also  have that,
$|\vec{Q}_m'^2|=2 $.
Since $I$ is given by, eq.(\ref{valnonsusyexI}), we see that condition eq.(\ref{two})
 is now met and the final
set of charges are in the Cardy limit.

To obtain an example with all final charges which are non-zero being much  bigger than unity
we can scale the initial charges, so that $(\vec{Q}_e,\vec{Q}_m)\rightarrow
(\lambda \vec{Q}_e,\lambda \vec{Q}_m), \lambda \gg 1$, and now take,
\beq
\label{nclam}
p \gg \lambda.
\eeq

Another example is as follows.
We take,
\beqa
\label{ex2ns}
\vec{Q}_e & = & (q_0,-p^1,0, 0, \cdots, 0) \\
\vec{Q}_m & = & (0,0, p^2, p^2, 0 , \cdots, 0),
\eeqa
with
\beq
\label{cexns3}
|q_0| \sim |p^1|.
\eeq
This system is not in the Cardy limit.

Applying the
 $O(6,22)$ transformation which acts non-trivially only on the 4 dimensional sublattice
gives by the first two factors of $\cal{H}$ in eq.(\ref{inpro}) and has the form,
\beq
\label{elslns3}
B=\pmatrix{1 & 0 & 0 & 0 \cr
           1 & 1 & 1 & -1 \cr
           1 & 0 & 1 & 0 \cr
           -1 & 0 & 0 & 1 },
\eeq
gives the final charges,
\beqa
\label{fchargeqe}
\vec{Q}_e' & = & (q_0, q_0-p^1, q_0,-q_0,0 \cdots 0) \\
\vec{Q}_m' & = & (0,0, p^2,p^2, 0, \cdots).
\eeqa
As long as the condition,
\beq
\label{cnse4}
|q_0p^1|\gg 6 (p_1-q_0)^2 (p^2)^2
\eeq
is met this final configuration satisfies eq.(\ref{two}) and is in the Cardy limit.

\section{More Details on Changing the Charges}
Two results of relevance to section 3.4 will be derived here.

First, we show that an $SL(2,\mathbb Z)$ matrix of the form, eq.(\ref{newsl}),
can always be found where $a,b$ meet the conditions,
eq.(\ref{ornewab}).

The integers, $m,n$ are determined in terms of the value of $\alpha$ for the altered charges,
eq.(\ref{falpha}). These can be taken to be coprime. Thus an $SL(2,\mathbb Z)$ matrix can always be found
of the form,
\beq
\label{alsl}
A'=\pmatrix{a' & b' \cr m & n \cr }
\eeq
The integers, $a',b'$ satisfy the condition,
\beq
\label{intp}
{\rm det}(A)= a'n-b'm = 1.
\eeq
From here it follows that,
\beq
\label{intp2}
\left[{a'\over m}\right]=\left[{b'\over n}\right]
\eeq
where  $[{a'\over m}]$ denotes the integer part of $|{a'\over m}|$, and similarly for
$[{b'\over n}]$.
Now, the allowed values of  integers, $a',b'$, which satisfy eq.(\ref{intp}) are not unique.
One can see that  if $a', b'$ satisfy eq.(\ref{intp}) then so do,
\beqa
\label{newab}
 a & = & a'-\left[{a'\over m}\right] m \\
b & = & b'-\left[{a'\over m}\right] n
\eeqa
From eq.(\ref{intp2}) it follows that the relations in eq.(\ref{ornewab}) are valid.
The resulting  $SL(2,\mathbb Z)$ transformation  is then given in eq.(\ref{newsl}).

Next we show that starting with the charges, eq.(\ref{faltch}), and applying the
$SL(2,\mathbb Z)$ transformation, eq.(\ref{newsl}), followed by an $O(6,22, \mathbb Z)$
transformation, gives rise to the charges, eq.(\ref{finalcharge}).
The $SL(2,\mathbb Z)$ transformation acting on eq.(\ref{faltch}) gives the charges,
\beqa
\label{intch}
\vec{\hat{Q}}_e=(a\tilde{q}_0,-a\tilde{p}^1, b\tilde{p}^i, 0, 0, 1,0) \cr
\vec{\hat{Q}}_m=(m\tilde{q}_0,-m\tilde{p}^1,n\tilde{p}^i,1,0,0,0).
\eeqa

Next, we determine the $O(6,22,\mathbb Z)$ transformation.
Consider a four dimensional subspace of the charge lattice,
where the metric, eq.(\ref{metric}), is,
$\cal{H} \oplus \cal{H}$. The following matrix is an element of $O(2,2,\mathbb Z)$,
\beq
\label{mato}
\pmatrix{1 & 0 & 0 & 0 \cr 0 & 1 & q & 0 \cr
         0 & 0 & 1 & 0 \cr -q & 0 & 0 & 1},
\eeq
for any $q\in \mathbb{Z}$.
Now starting with the charges, eq.(\ref{intch}),
consider such a transformation, with $q=m\tilde{p}^1$,  acting on the charges lying in the
first Hyperbolic subspace and the second last Hyperbolic
 subspace, as defined in eq.(\ref{metric}). And next such a transformation, with $q=(a \tilde{p}^1+1)$,
 acting on the
charges in the first Hyperbolic subspace and the last Hyperbolic subspace, as defined in eq.(\ref{metric}).
This takes the charges in  eq.(\ref{intch}) to their final values in eq.(\ref{finalcharge}).

\section{Some more details on the Isometry Analysis of Section 5 }

In this section we will derive all the isometries preserved by the metric eq.(\ref{newmc}).
The Killing vectors must satisfy the conditions  given by eq.(\ref{kcond}).
The $(\theta_1, \theta_1), (\theta_2,\theta_2), (\theta_1,\theta_2)$ components of this
equation take the form,
\beqa
\label{firstthree1}
\partial_{\theta_1}\xi^{\theta_1} & = & 0 \cr
\partial_{\theta_2}\xi^{\theta_2} & = & 0 \cr
\partial_{\theta_1}\xi^{\theta_2}+\partial_{\theta_2}\xi^{\theta_1} & = & 0.
\eeqa
The $(\phi_1,\phi_1), (\phi_2,\phi_2), (\phi_1,\phi_2),$ components are,
\beqa
\label{nextthree1}
im_1\xi_{\phi_1} + \alpha^2 \xi^{\theta_1}\sin\theta_1\cos\theta_1 & = & 0 \cr
i m_2 \xi_{\phi_2} + \beta^2 \xi^{\theta_2} \sin\theta_2\cos\theta_2 & = & 0 \cr
im_1 \xi_{\phi_2}+im_2\xi_{\phi_1}-\sin\theta_1\cos\theta_2\xi^{\theta_1}
-\sin\theta_2\cos\theta_1\xi^{\theta_2} & = & 0
\eeqa
The $(\psi,\psi), (\psi,\phi_1), (\psi,\phi_2),$ components are,
\beqa
\label{nextnextthree1}
im_3 \xi_\psi & = & 0 \cr
im_1\xi_{\psi}+ im_3\xi_{\phi_1}-\sin\theta_1 \xi^{\theta_1} & = & 0 \cr
im_2\xi_{\psi}+ im_3\xi_{\phi_2}-\sin\theta_2 \xi^{\theta_2} & = & 0
\eeqa

The $(\theta_1,\phi_1), (\theta_2,\phi_2), (\theta_1,\phi_2), (\theta_2\phi_1),$
components are,
\beqa
\label{nnnfour1}
\partial_{\theta_1}\xi^\gamma g_{\gamma \phi_1}+im_1\xi^{\theta_1} & = & 0 \cr
\partial_{\theta_2}\xi^\gamma g_{\gamma \phi_2}+im_2\xi^{\theta_2} & = & 0 \cr
\partial_{\theta_1}\xi^\gamma g_{\gamma \phi_2}+im_2\xi^{\theta_1} & = & 0 \cr
\partial_{\theta_2}\xi^\gamma g_{\gamma \phi_1}+im_1\xi^{\theta_2} & = & 0
\eeqa
Finally the $(\theta_1,\psi), (\theta_2,\psi),$ components are,
\beqa
\label{lasttwo1}
\partial_{\theta_1}\xi^\gamma g_{\gamma\psi}+im_3\xi^\theta_1 & = & 0 \cr
\partial_{\theta_2}\xi^\gamma g_{\gamma\psi}+im_3\xi^\theta_2 & = & 0
\eeqa

Setting $m_1=m_2=m_3=0$ we  have from the $(\psi,\phi_1)$ and $(\psi,\phi_2)$
components   that, $\xi^\theta_1=\xi^\theta_2=0$. It then follows from the
remaining equations that there are only three Killing vectors of this type.
These are, $\partial_\psi, \partial_{\phi_1},\partial_{\phi_2}$, which have already been
identified above.

Next setting $m_1\ne 0, m_2\ne 0, m_3 \ne 0$ we have, from the equation for
$(\psi, \psi)$, $(\phi_1,\phi_1)$ and $(\psi,\phi_1)$ components that,
\beq
\label{cisoa1}
-{\alpha^2\cos\theta_1 \over m_1} \xi^\theta_1={1\over m_3}\xi^\theta_1,
\eeq
from which we conclude that
\beq
\label{cisob1}
\xi^\theta_1=0.
\eeq
Similarly we learn that $\xi^\theta_2=0$. From the $(\phi^1,\phi^1), (\phi^2,\phi^2),
(\psi,\psi),$ components it then follows that,
\beq
\label{cisoc1}
\xi_\mu=0 \ \  \forall \  \mu,
\eeq
leading to the conclusion that there is no Killing vector of this type.

We will now set $m_1=m_2=0$ and $m_3\neq 0$. The $(\phi_1,\phi_1)$ and $(\phi_2,\phi_2)$ components
give, respectively, $\xi^{\theta_1}=0$ and $\xi^{\theta_2}=0$. The $(\psi,\gamma)$ components for
$\gamma = \psi,\phi_1$ and $\phi_2$ give $\xi_\psi=0,\xi_{\phi_1} = 0$ and $\xi_{\phi_2}=0$ respectively.
Thus we have no killing vector with $m_1=m_2=0$ and $m_3\neq 0$.

Let us now set $m_2 = m_3 = 0$ and $m_1 \neq 0$. Considering the $(\phi_2,\phi_2)$ component, we get
$\xi^{\theta_2}=0$. From the $(\phi_1,\phi_1), (\phi_1,\phi_2)$ and $(\psi,\phi_1)$ components we get,
\beqa
\xi_{\phi_1} &=& - \left(\frac{\xi^{\theta_1}}{im_1}\right) \alpha^2 \sin\theta_1\cos\theta_1 \cr
\xi_{\phi_2} &=& \left(\frac{\xi^{\theta_1}}{im_1}\right) \sin\theta_1 \cos\theta_2 \cr
\xi_\psi &=& \left(\frac{\xi^{\theta_1}}{im_1}\right) \sin\theta_1 ~.
\eeqa
The contravariant components of $\xi$ can be shown to be
\beqa
\xi^{\phi_1} &=& -\left(\frac{\xi^{\theta_1}}{im_1}\right) \cot\theta_1 \cr
\xi^\psi &=&  \left(\frac{\xi^{\theta_1}}{im_1}\right) {\rm cosec}\ \theta_1 ~,
\eeqa
and $\xi^{\phi_2} = 0$. We still have to satisfy the remaining nontrivial equations. The $(\theta_1,\phi_1)$
component of the killing equation
\beqa
\partial_{\theta_1}\xi^{\phi_1} g_{\phi_1\phi_1} + \partial_{\theta_1}\xi^\psi g_{\psi\phi_1}
+ i m_1 \xi^{\theta_1} = 0  ~,
\eeqa
gives
\beq
- \frac{1}{m_1} (1+\alpha^2) + m_1 = 0 ~.
\eeq
Thus we must have
\beq
m_1 = \pm \sqrt{1+\alpha^2} ~.
\eeq
It is straightforward to check that the $(\theta_1,\phi_2)$ and $(\theta_1,\psi)$ components of the killing
equation are satisfied. All other components are satisfied trivially provided $\xi^{\theta_1}$ is independent of $\theta_1, \theta_2$. As a result we get two linearly independent killing vectors corresponding to the two roots
of $m_1$:
\beqa
\xi_1 &=& e^{i\sqrt{1+\alpha^2}\phi^1} \left(\partial_{\theta_1}
+ \frac{i}{\sqrt{1+\alpha^2}} \cot\theta_1 \partial_{\phi_1} - \frac{i}{\sqrt{1+\alpha^2}} \rm{cosec}\ \theta_1
\partial_\psi\right)~, \cr
\xi_2 &=& \xi_1^* ~.
\eeqa
In a similar way we can obtain two more linearly independent killing vectors upon setting $m_1=m_3=0$ and
$m_2\neq 0$. We find
\beqa
\xi_3 &=& e^{i\sqrt{1+\beta^2}\phi^1} \left(\partial_{\theta_2}
+ \frac{i}{\sqrt{1+\beta^2}} \cot\theta_2 \partial_{\phi_2} - \frac{i}{\sqrt{1+\beta^2}} \rm{cosec}\ \theta_2
\partial_\psi\right)~, \cr
\xi_4 &=& \xi_3^* ~.
\eeqa

Let us now set $m_1\neq 0,m_2\neq 0$ and $m_3=0$. The $(\psi,\phi_1)$ and $(\psi,\phi_2)$ components together
gives
\beqa
i m_1 \xi_\psi - \sin\theta_1 \xi^{\theta_1} &=& 0 \cr
i m_2 \xi_\psi - \sin\theta_2 \xi^{\theta_2} &=& 0 ~.
\eeqa
Eliminating $\xi_\psi$ from the above two equations, we find
\beq
\frac{\xi^{\theta_1}}{\xi^{\theta_2}} = \frac{m_1 \sin\theta_2}{m_2 \sin\theta_1} ~.
\eeq
Since $\xi^{\theta_1}$ is independent of $\theta_1$ and $\xi^{\theta_2}$ is independent of $\theta_2$, the
above equation can be met only if $\xi^{\theta_1}$ is proportional to $\sin\theta_2$ and vice versa. From
$\partial_{\theta_1}\xi^{\theta_2} + \partial_{\theta_2}\xi^{\theta_1} = 0$ we find $\partial_{\theta_1}
\partial_{\theta_2}\xi^{\theta_2} = 0$, indicating the proportionality constants must be zero. From the
above discussion, we get $\xi^{\theta_1} = \xi^{\theta_2} = \xi_\psi = 0$. It is now easy to see from the
$(\phi_1,\phi_1)$ and $(\phi_2,\phi_2)$ components of the killing equation that $\xi_{\phi_1}=\xi_{\phi_2}=0$.
And hence we don't have any killing vector for the above choice of $m_1,m_2,m_3$. In a similar manner, we
can show hat we don't have any nontrivial solution to the killing equations when $m_1\neq 0, m_3\neq 0$
and $m_2=0$ as well as when $m_2\neq 0, m_3\neq 0$ and $m_1=0$.

In summary, the metric, eq.(\ref{newmc}), has seven Killing vectors, given in eq.(\ref{kvectors}).

\section{General canonical form of charge vector in $\Gamma^{6,6}$}
We start with a charge vector $\vec{Q} \in \Gamma^{2,2}$, where $\Gamma^{2,2} = \cal{H} \oplus
\cal{H}$,  is the
$4$-dimensional lattice made out of two 2-dimensional Hyperbolic  lattices, $\cal{H}$.
In components, $\vec{Q}$ takes the form,
\beq
\label{aecfq}
\vec{Q}=(a,-b,c,d).
\eeq
The lattice, $\Gamma^{2,2}$, is  invariant under the  action of $O(2,2,\mathbb Z)$.
We show that using an   $SL(2, \mathbb Z) \times SL(2, \mathbb Z) \in O(2,2,Z)$  the
vector, $\vec{Q}$,
 can be brought to the  form,
\beq
\label{aecfq2}
\vec{Q}=(gcd(\vec{Q}), {\vec{Q}^2 \over gcd(\vec{Q})}, 0, 0),
\eeq
where,
\beq
\label{defgcdae}
gcd(\vec{Q})=gcd(a,b,c,d),
\eeq
and
\beq
\label{defnormae}
\vec{Q}^2=\vec{Q} \cdot \vec{Q}.
\eeq
Note that the only non-vanishing components in eq.(\ref{aecfq2}) lie in the first $\cal{H}$
sublattice.

It is useful for this purpose  to represent $\vec{Q}$ as a $2\times 2$ matrix,
\beq
\label{matformq}
Q=\pmatrix{a & -b \cr c & d \cr}.
\eeq
The first $SL(2, \mathbb Z)$, which we denote as $SL(2,\mathbb Z)_T$, acts on the left and
performs row operations, while the second  $SL(2, \mathbb Z)$,
 which we denote as $SL(2,\mathbb Z)_U$, acts on the right and carries out column operations.
If $A \in SL(2, \mathbb Z)_T, B \in  SL(2, \mathbb Z)_U$, then under their action,
\beq
\label{matac}
Q\rightarrow A Q B.
\eeq
Note that $\vec{Q}^2 = det(Q)$. We will show that $A,B$ can be found which bring $Q$ to the
 form,
\beq
\label{finmatac}
Q=\pmatrix{gcd(\vec{Q}) & 0 \cr 0  & {det(Q) \over gcd(\vec{Q})} \cr }
\eeq
This is equivalent to $\vec{Q}$ taking the form, eq.(\ref{aecfq2}).

It is enough to prove this  result for the case when $gcd(\vec{Q})=1$, in which case,
eq.(\ref{finmatac}) becomes,
\beq
\label{gcof}
Q=\pmatrix{1 & 0 \cr 0  & det(Q) \cr }.
\eeq
 The more general result, eq.(\ref{finmatac}), then follows, by considering the vector,
${1\over gcd(\vec{Q})} \vec{Q}$, which has unit value for its gcd.
In the discussion below we will sometimes use to the notation,
\beq
\label{notgcd}
gcd(Q)\equiv gcd(\vec{Q}) =gcd(a,b,c,d).
\eeq

The proof is as follows.
Given any 2 integers, Euclid gives us an algorithm to arrive at their gcd in the
following fashion. Subtract the smaller of the 2 numbers from the
larger
and then if the result is still larger than the
smaller number continue this operation till the result becomes otherwise.
Then start subtracting the new smaller number from the new larger number and continue
this set of steps till one of the numbers becomes  zero at which point the other number
is the gcd.
If the two integers are $a,c$,  the two elements of the first column of  matrix, $Q$,
eq.(\ref{matformq}),  then this sequence of operations can be implemented by an element of
$Sl(2,\mathbb Z)_T$ which acts on the left and carries out row operations.
The resulting form of $Q$ is,
\beq
\label{formqa}
Q=\pmatrix{a' & b' \cr 0 & d'},
\eeq
where $a'=gcd(a,c)$.
Note that  $gcd(Q)$ is preserved by this operation. Since $gcd(Q)=1$, to begin with, we
learn that,
\beq
\label{gcdcae}
gcd(a',b',d')=1.
\eeq

Now we come to the crucial step. Let $\{p_1, \cdots p_r\}$, be the set of distinct primes
which divide $d'$ but do not divide $b'$. Let $m=\Pi p_i$, be the product of all these primes.
One can show that the two numbers, $d'$, and, $a'm +b'$, are coprime.
Let $p'$ be a prime that divides $d'$, then if it does not divide $b'$ it must divide $m$
(by construction) and thus cannot divide $a'm+b'$. If on the other hand $p'$ divides $b'$,
it cannot divide $m$ (again by construction)  and also
it cannot divide $a'$ (since eq.(\ref{gcdcae}) is valid), and therefore  $p'$ cannot divide
$a'm+b'$.  Thus, we learn that $gcd(d', a'm + b')=1$ and these two numbers are coprime.

We use this result to bring $Q$, eq.(\ref{formqa}), to the form,  eq.(\ref{gcof}).
First, an $SL(2,\mathbb Z)_U$ transformation can be carried out,
\beq
\label{changeae1}
Q \rightarrow Q \pmatrix{1 & m' \cr 0 & 1 \cr} = \pmatrix{a' & a'm +b' \cr 0 & d' \cr}.
\eeq
Since $gcd(a'm+b', d')=1$,   we can use Euclid's algorithm as in the discussion
above to  now find  an $SL(2, \mathbb Z)_U$ transformation which bring $Q$ to the form,
\beq
\label{changeae2}
Q=\pmatrix{a'' & 1 \cr c'' &  d'' \cr }.
\eeq
Next, further $SL(2, \mathbb Z)_T \times SL(2, \mathbb Z)_U$ tranformations can be carried out
to subtract the second column from the first $a''$ times, and the first row from the second
$d''$ times. This followed by a row- column interchange operation gives $Q$ in the form,
$\pmatrix{1 & 0 \cr 0  & u\cr }$.
Since these operations preserve the determinant, we learn that $u=det(Q)$, leading to
eq.(\ref{gcof}).

We end by making a few points.  First,  note that this argument holds for space-like, time-like
and null charge vectors, $Q$. Second,
it follows from our analysis that there are  two independent
 invariants for $SL(2,\mathbb Z) \times
SL(2,\mathbb Z)$. These are  $det(Q)$ and $gcd(Q)$. Of these $det(Q)$ is an invariant
of the continuous group, while $gcd(Q)$ is a discrete invariant.
Third, if instead of $\Gamma^{2,2}$ we start with a lattice
which is the direct sum of more than two copies of $\cal{H}$, a similar argument can be
used sequentially on the first two $\cal{H}$ sublattices, then the first
and third $\cal{H}$ sublattices etc, to finally bring the charge vector to  the form,
\beq
\label{finalformqe}
\vec{Q}=(gcd(\vec{Q}), {\vec{Q}^2\over gcd(\vec{Q})}, 0 ,0 , \cdots, 0, 0).
\eeq
In particular this is true for $\Gamma^{6,6}$ which consists of six copies of $\cal{H}$.
Finally, if there are two charge vectors, $\vec{Q}_e, \vec{Q}_m$, then  the above argument
can be used to put one of them, say $\vec{Q}_e$, in the form, eq.(\ref{finalformqe}).
Further transformations which act trivially on the first Hyperbolic sublattice will keep
$\vec{Q}_e$ invariant. Using these transformations $\vec{Q}_m$ can now  be brought to the form,
\beq
\label{finalformqm}
\vec{Q}_m=(\alpha, \beta, \gamma, \delta, 0 ,0 \cdots, 0 ,0),
\eeq
so that only the components in the first two Hyperbolic sublattices are non-vanishing.
These results apply in general to the cases when
$\vec{Q}_e^2, \vec{Q}_m^2$,
 have space-like, time-like or null norms.

\bibliographystyle{JHEP}

\bibliography{draft1111}

\end{document}